\documentclass[sn-mathphys]{sn-jnl}

\jyear{2022}%

\usepackage{latexsym}
\usepackage{graphics}
\begin{document}
\title[Electromagnetic interaction of low transverse momentum electrons and positrons]{Electromagnetic interaction of low transverse momentum electrons and positrons
\\with heavy nuclei in ultra-peripheral ultra-relativistic heavy-ion collisions}



\author*[1]{\fnm{Katarzyna} \sur{Mazurek}}\email{Katarzyna.Mazurek@ifj.edu.pl}

\author[1]{\fnm{Mariola} \sur{K\l{}usek-Gawenda}}\email{Mariola.Klusek-Gawenda@ifj.edu.pl}
\equalcont{These authors contributed equally to this work.}

\author[1,2]{\fnm{Antoni} \sur{Szczurek}}\email{Antoni.Szczurek@ifj.edu.pl}
\equalcont{These authors contributed equally to this work.}

\affil*[1]{\orgdiv{}, \orgname{Institute of Nuclear Physics Polish Academy of Sciences}, \orgaddress{\street{Radzikowskiego 152}, 
\city{Krak\'ow}, \postcode{PL-31342}, \state{}, \country{Poland}}}

\affil[2]{\orgdiv{College of Natural Sciences, Institute of Physics}, \orgname{University of Rzesz\'ow}, \orgaddress{\street{Pigonia 1}, 
\city{Rzesz\'ow}, \postcode{PL-35310}, \state{}, \country{Poland}}}


%

\abstract{
The photon-photon interaction in ultra-relativistic heavy-ion collision
is a source of the $(e^+,e^-)$ pairs. 
The photon-photon fusion leads the lepton creation 
in the broad configuration space around ''collision'' point.
Those created close to heavy nuclei may undergo strong electromagnetic interaction with them. 
The impact parameter space distribution of electrons and positrons are calculated within 
the b-space EPA model of such collisions. 
The evolution due to the electromagnetic final state interaction (FSI) of $(e^+,e^-)$
with nuclei is studied, and the distortion of rapidity and transverse 
momentum distributions are shown. Part of the analysis is independent of the production model.
We show first exploratory results for the reaction Pb+Pb at 
$\sqrt{s_{NN}}=$ 17.3~GeV (SPS) and 200~GeV (RHIC) energies.
We provide results for selected creation points and when integrating
over their position as dictated by the $b$-space EPA model.
We observe a strong influence at low transverse momenta, so far not
measured regions of the phase space. In particular, we predict 
a possible sizeable accumulation of electrons with rapidities close 
to the beam rapidity.
The EM effects lead to asymmetry in the production of electrons and 
positrons. 
} 
%


%


\maketitle


\section{Introduction}
\label{Sect.I}

Heavy-ion collisions are excellent factory for producing 
both elementary and composed particles as well as for studying their 
properties and production mechanism. Since many years 
efforts of theorists and experimentalists were focused on the investigation 
of time-space evolution of the quark-gluon plasma (QGP) and production of
different species of particles, primarily hadrons 
(pions, kaons, nucleons, hyperons, etc.) emitted in the collision. 
At high energies, the velocities of such beam nuclei are close to light 
velocity thus they are often called ultra-relativistic 
velocities (URV). 

Central collisions are the most interesting in the context of
the QGP studies. Plasma is, of course, also produced in more peripheral 
collisions.
In peripheral collisions, the so-called spectators are relatively large 
and have large moving charge. It was realized relatively late that 
this charge generates strong quickly changing electromagnetic fields 
that can influence the trajectories and some observables 
for charged particles.

Such effects were investigated in previous studies of one of the present
authors \cite{Rybicki:2006qm,Rybicki:2013qla}.
On one hand side the EM effects strongly modify the Feynman $x_F$ spectra
of low-$p_T$ pions, creating a dip for $\pi^+$ and
an enhancement for $\pi^-$ at $x_F \approx \frac{m_{\pi}}{m_N}$.
In \cite{Rybicki:2006qm} a formalism of charged meson evolution
in the EM field (electric and magnetic) of fast moving nuclei was
developed. Later on the spectacular effects were confronted with 
the SPS data \cite{Rybicki:2009zz} confirming the theoretical predictions.
The investigation was done for $^{208}$Pb+$^{208}$Pb at 158~GeV/nucleon 
energy ($\sqrt{{s}_{NN}} =$ 17.3~GeV) at 
CERN Super Proton Synchrotron (SPS) \cite{Schlagheck:1999aq}.
In \cite{Rybicki:2013qla} the influence of the EM fields
on azimuthal flow parameters ($v_n$) was studied and confronted in
\cite{Rybicki:2014rna} with the RHIC data. It was found that the EM field
leads to a split of the directed flow for opposite e.g.
charges. In the initial calculation, a simple model of single initial
creation point of pions was assumed for simplicity.
More recently, such a calculation was further developed by taking 
into account also the time-space evolution of the fireball, treated 
as a set of firestreaks \cite{Ozvenchuk:2019cve}. 
The distortions of the $\pi^+$ and $\pi^-$ distributions allow to
discuss the electromagnetic effects of the spectator and charged pions 
in URV collisions of heavy ions and nicely explain the experimental data. 
This study was done for non-central collisions where the remaining 
after collision object, called 'spectator', loses only a small part 
of nucleons of the original beam/target nucleus.
 
The ultra-peripheral collisions of heavy ions (e.g.~$^{208}$Pb +$^{208}$Pb) at 
ultra-relativistic energies ($\sqrt{s_{NN}}\ge 5~$GeV) \cite{Klusek-Gawenda:2016suk}
allow to produce particles in a broad region of impact parameter
space, even far from ``colliding'' nuclei.
The nuclei passing near one to each other with ultrarelativistic energies 
are a source of virtual photons that can collide producing
e.g. a pair of leptons.
In real current experiments (RHIC, LHC), the luminosity is big enough to observe
e.g. $AA \rightarrow AA\rho^0$ and $AA \rightarrow AAe^+e^-$, 
$AA \rightarrow AA\mu^+\mu^-$ processes. 
One of the most interesting phenomena is multiple interaction 
\cite{Klusek-Gawenda:2016suk,vanHameren:2017krz}
which may lead to the production of more than one lepton pair.

The studies on the creation of positron-electron pairs started in early 
1930'ties with prediction of positron by Dirac \cite{Dirac:1930}, experimental discovery of positron by Anderson \cite{Anderson} and works
of Breit and Wheeler \cite{Breit:1934zz}, where they calculated 
the cross section for the production of such pairs in the collisions of two light quanta. 
The production of $e^+e^-$ pairs in nucleus-nucleus collisions, goes back to the work on Bethe, Heitler \cite{BH1934} and Landau, Lifschitz \cite{LL1934} in 1934 .
It was Williams who realized \cite{Williams:1935} that the 
production of $e^+ e^-$ pairs is enhanced in the vicinity of the atomic
nucleus. 
An overview of the theoretical investigation of 
the $e^+ e^-$ pairs creation in historical context was presented 
by e.g. Hubbell \cite{Hubbel:2006} 
and the detailed discussion about this process in physics and
astrophysics was written by Ruffini et al. \cite{Ruffini:2009hg}.

The early analyses were done in the momentum space and therefore
did not include all details in the impact parameter space.
An example of the calculation where such details are taken into
account can be found e.g. in 
\cite{Klusek-Gawenda:2010vqb,Klusek-Gawenda:2016suk}.
In ultra-peripheral collision (UPC), the nuclei do not collide
without loosing, in principle, any nucleon.
However, the electromagnetic final state interaction (FSI) induced by fast moving
nuclei may cause excitation of the nuclei and subsequent emission 
of different particles, in particular neutrons \cite{Klusek-Gawenda:2013ema}
that can be measured both at RHIC and at the LHC.
Moreover the UPC are responsible not only 
for Coulomb excitation of the spectator but also for 
the multiple scattering and production of more than one dielectron
pair \cite{vanHameren:2017krz}. 
With large transverse momentum cut typical at RHIC and the LHC the effect
is not dramatic.

New experimental results have shown that the $b$-space EPA may be not sufficient
to describe some observables \cite{STAR1}, such as $p_{t,pair}$
distribution.
Very recently experimental results for the semi-central collisions
controlled by centrality measurement have shown a strong enhancement at 
$p_{t,pair} \sim$ 0 (STAR) \cite{STAR2} and a bump at small acoplanarity
(ATLAS) \cite{ATLAS1,ATLAS2}. Inclusion of transverse momentum dependent photon distributions
improved the situation \cite{k_t-factorization} but did not give 
the observed experimentally dependence on impact parameter.
Recently several works addressed the issue of impact parameter -- 
transverse momentum correlations 
\cite{LZZ2019,ZBTX2020,KMXY2020,WPW2021}
for different interesting observables.
A full treatment is a bit difficult \cite{Wigner} but gives very good
description of the semi-central collision data.

Can the strong EM fields generated at high energies modify 
the electron/positron distributions?
No visible effect was observed for electrons with $p_T >$ 1~GeV
as discussed in \cite{vanHameren:2017krz}, where the ALICE distributions
were confronted with the $b$-space equivalent photon approximation (EPA)
model.
But the electromagnetic effect is expected rather at 
low transverse momenta.
According to our knowledge this topic was not discussed 
in the literature.
As the spectators, which in ultra-peripheral collisions are almost 
identical to colliding nuclei, are charged,
they can interact electromagnetically with electrons and positrons as 
it was in the case of pions. 
Similar effects as observed for pions may be expected also for charged 
leptons. The motion of particles in EM field depends not only on their 
charge but also on their mass. Thus the distortions of $e^+/e^-$
distributions should be different than those for $\pi^+/\pi^-$ distributions. 
Also the mechanism of production is completely different. 
In contrast to pion production, where the emission site is well localized,
the electron-positron pairs produced by photon-photon fusion can 
be produced in a broad configuration space around the ``collision''
point - point of the closest approach of nuclei. 
A pedagogical illustration of the impact parameter dependence can 
be found e.g. in \cite{Klusek-Gawenda:2010vqb}.

From one side, the previous works \cite{Ozvenchuk:2019cve} and reference 
therein, were connected with the electromagnetic effects caused by 
the emission of the pions from the fireball region. 
From the other side, the model considered \cite{Klusek-Gawenda:2010vqb,Klusek-Gawenda:2018zfz,Klusek-Gawenda:2020eja} can correctly estimate 
the localization in the impact parameter space.
The present study will be focused on electromagnetic 
interaction between electrons/positrons with highly positively charged nuclei.

Our approach consists of two steps. First, the $e^+ e^-$ distributions 
are calculated within EPA in terms of initial distributions in a given space 
point and at a given initial rapidity and transverse momentum.
Secondly, the space-time evolution of leptons in the electromagnetic fields
of fast moving nuclei with URV is performed by solving relativistic 
equation of motion \cite{Rybicki:2011zz}.

In Section \ref{Sect.II} the details of the calculation of 
the differential cross section of $e^+ e^-$ production will be
presented.
We will not discuss in detail equation of motion which was presented
e.g. in \cite{Rybicki:2011zz}. 
The results of the evolution of electrons/positrons in
the EM field of ``colliding'' nuclei are presented in Section \ref{Sect.III}.

\section{A comment on quantum multiple-photon exchange effects}

The electromagnetic interaction of charged leptons with nuclei was
discussed in many contexts in the literature.
Very similar to the considered here process is $\gamma A \to l^+ l^- A$ 
reaction.
The respective Coulomb effect was discussed e.g. in 
\cite{BM1954,IM1998,Tuchin2009,SZZZ2020}.

The history started with the seminal work of Bethe and Maximon
\cite{BM1954}. They discussed the Coulomb effects for bremsstrahlung and dilepton pair production.
Tuchin considered \cite{Tuchin2009} multiple $t-$channel photon exchanges 
within the Glauber model.
Ivanov and Melnikov \cite{IM1998} presented impact representation of 
the amplitude for $e^+ e^-$ photoproduction in the Coulomb field of the nucleus.
They wrote leading amplitudes for N $t$-channel exchanges and 
showed that in their approximations the amplitudes can be resummed. 
They calculated corrections to the total $\gamma A \to l^+ l^- A$
cross section which is of course much simpler than for differential
distributions (not discussed there). 

Very recently the Shandong group \cite{SZZZ2020} discussed how 
to modify the TMD photon distributions in the nucleus due to strong Coulomb 
field of the nucleus. This approach may be relevant for $A(e,e'l^+ l^-)A$
at EIC and EicC.
Also, this approach is not very useful in our case,
where we need rather impact parameter formulation than momentum space
formulation.

Very closely related are Coulomb effects in
$\mu + A \to \mu + A + e^+ e^-$ \cite{IKSS1998,LMS2004,SR2018}.
In \cite{LMS2004} the authors discuss distributions
in $x = \epsilon_-/\omega$, where $\epsilon_-$ is electron energy 
and $\omega$ is initial photon energy.
In a recent paper \cite{SR2018} the authors discuss distributions in
$x = \epsilon_+/\omega$, where $\epsilon_+$ is positron energy
and distribution in muon energy loss through pair production
relevant for muons propagating in the atmosphere.

\begin{figure}[!hbt]
\resizebox{0.49\textwidth}{!}{%
     \includegraphics{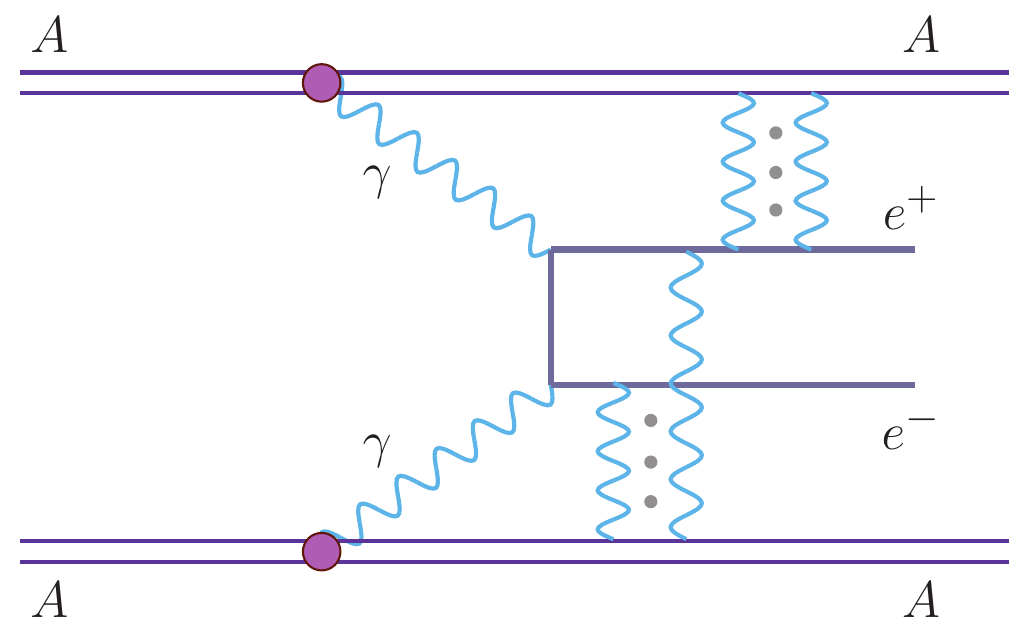}
     }
     \caption{An example of diagrams relevant for multi-photon exchanges.}
     \label{AA_AAepem}
\end{figure}

The $A A \to A A l^+ l^-$ reaction was also discussed in the literature
\cite{SW1998,SW1999,ERSG1999,ISS1999,BGLP2001,LMS2002,ML1998,B2003,BGKN2002,BGKN2004}.
In Fig.\ref{AA_AAepem} we show an example
of the amplitude for a few photon exchanges. It becomes obvious
that resummation with exact kinematics and extended charges of nuclei 
is not easy.

The  Massachusetts group discussed time-dependent two-center Dirac 
equation with Coulomb interaction \cite{SW1998,SW1999}.
The papers have rather technical character.
Neither quantitative predictions were given nor differential
distributions were discussed.

The Frankfurt group \cite{ERSG1999} discussed how to modify the
impact parameter dependent photon distributions in the strong
electromagnetic field of the nuclei. As stated by the authors
their approach is not adequate when lepton velocities are similar
to the velocity of one of the nuclei. In the present paper we wish to
discuss exactly this region of the phase space.
The classical approach performed in \cite{Rybicki:2006qm,Ozvenchuk:2019cve} for charged pion 
production in heavy ion collisions suggests that this is the region
where spectacular effects may occur.

The authors of \cite{BGKN2004} reported an explicit calculation of the
contribution of exchange of two photons from one nucleus and two photons from the
other nucleus. Their results suggest that the familiar eikonalization of
Coulomb distortions breaks down. Some relations to 
$n \gamma + m \gamma \to l^+ l^-$ were discussed there.
A full formalism is not ready for practical calculations.

The short summary is intended to show that there is no ready formalism
which can be used to calculate Coulomb modifications of 
differential distributions of individual, positively or negatively 
charged, leptons. Therefore in the following we suggest to use
rather existing classical approach which seems much easier to apply
for our purpose. We shall return to the discussion of this approach
in the following sections.


\section{Lepton pair production, equivalent photon approximation}
\label{Sect.II}
%

The particles originated from photon-photon collisions can be created
in full space around excited nucleus thus first of all the geometry of 
the reaction should be defined.  

In the present study, the ultra-peripheral collisions (UPC) are
investigated in the reaction plane ($b_x,b_y$) which are perpendicular 
to the beam axis taken as $z$-direction.

\begin{figure}[!hbt]
\resizebox{0.49\textwidth}{!}{%
     \includegraphics{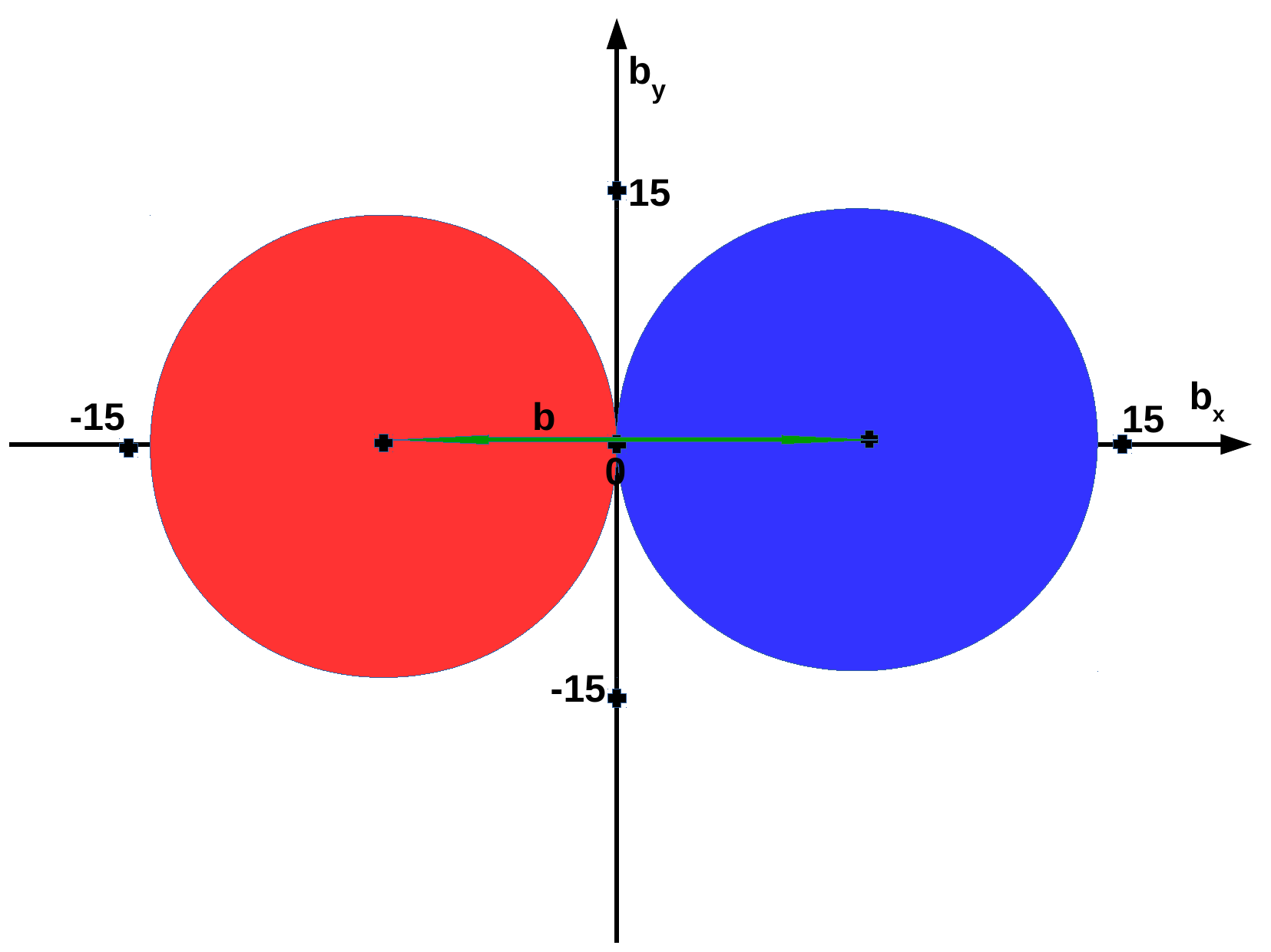}
     }
     \caption{ The impact parameter space and the
       five selected points ($b_x,b_y$):
     (0, 0), ($\pm$15~fm,0), (0,$\pm$15~fm), for which 
      the distribution in rapidity and transverse momentum will be
      compared latter on in the text, shown in the CM rest frame.
      }
     \label{fig01}
\end{figure}

The collision point ($b_x=0,b_y=0$) is time-independent center of mass (CM)
of the reaction as shown in Fig.~\ref{fig01}. The impact parameter
is fixed as double radius of each (identical) Pb nucleus b=(13.95~fm, 14.05~fm). 
For comparison we will show results also for b=(49.95~fm, 50.05~fm).

Four characteristic points ($\pm$15~fm, 0), (0, $\pm$15~fm), which are discussed later are also 
marked in the figure. In the present paper 
we shall show results for these initial emission points for 
illustrating the effect of evolution of electrons/positrons in 
the EM field of nuclei.

Here we wish to present some preliminary results for selected values
of impact parameter.  
Till recently the $b$-space EPA \cite{BF90} was state of the art \cite{Bertulani:2005ru,BGKN2009,STARlight}
for calculating dilepton production cross section,
although a slightly different approach, with impact parameter - lepton
momentum correlations, was available \cite{Vidovic1992,HTB1995}.
Usually the exclusive dilepton production was estimated by using 
the monopole charge form factor which allows to reproduce correctly 
the total cross section. 
The differential cross sections are more sensitive to
details, thus realistic charge form factor (Fourier transform of the
charge distribution) has to be employed.
\footnote{Double scattering production of positron-electron pairs using 
the realistic charge form factor has been discussed 
in \cite{Klusek-Gawenda:2016suk} and \cite{vanHameren:2017krz}.} 
In the $b$-space EPA (see \cite{BF90}) the total cross section for $l^+l^-$
pair production via photon-photon fusion reads
\begin{eqnarray}
\sigma &=& \int \frac{d\omega_1}{\omega_1} \int \frac{d\omega_2}{\omega_2} N\left( \omega_1, \textbf{b}_1 \right) N\left( \omega_2, \textbf{b}_2 \right) 
\Theta \left( b-R_1-R_2 \right)  \nonumber \\
&\times& 2\pi \int_{R_1}^{\infty} b_1 db_1 \int_{R_2}^{\infty} b_2 db_2 
\int_{0}^{2\pi} d \phi \, \sigma_{\gamma\gamma}\left(\omega_1,\omega_2\right)\,.
\end{eqnarray}
This can be written in the equivalent way as \cite{Klusek-Gawenda:2010vqb}
\begin{eqnarray}
&&\sigma_{AA \to AA e+e-}=  \nonumber \\
&&\int \frac{{\rm d} \sigma_{\gamma\gamma \to e^+e^-} \left(W_{\gamma\gamma}\right)}{{\rm d}\cos \theta} N\left( \omega_1, \textbf{b}_1 \right) N\left( \omega_2, \textbf{b}_2 \right) 
S^2_{abs}\left( \textbf{b} \right) \nonumber \\
&&\times 
2\pi b {\rm d} b {\rm
	d} \overline{b}_x \, {\rm d} \overline{b}_y 
 \frac{W_{\gamma\gamma}}{2}
{\rm d} W_{\gamma \gamma} \, {\rm d} Y_{e^+e^-} \,  {{\rm d}\cos \theta}  \;.
\label{EPA}
\end{eqnarray}
where $N(\omega_i,b_i)$ are photon fluxes, $W_{\gamma\gamma}=M_{e^+e^-}$
 is invariant mass and $Y_{e^+e^-}= (y_{e^+} + y_{e^-})/2$ is rapidity of 
the outgoing system and $\theta$ is the scattering angle in the $\gamma\gamma\rightarrow e^+e^-$ 
center-of mass system. The gap survival factor $S^2_{abs}$ assures that 
only ultra-peripheral reactions are considered.
%
\begin{figure}[!hbt]
\resizebox{0.49\textwidth}{!}{%
     \includegraphics{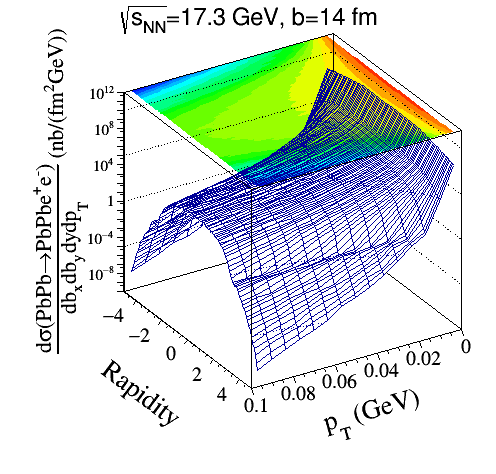}
     }
     \caption{The map of differential cross section in rapidity of
       electron or positron and lepton transverse momentum 
       for ($b_x,b_y$) = (0, 0) which will be called CM point for brevity.}
     \label{fig02a}
\end{figure}
%

The
$\overline{b_x}=(b_{1x}+b_{2x})/2$ and
$\overline{b_y}=(b_{1y}+b_{2y})/2$
quantities are particularly useful for our purposes.
We define $\vec{{\bar b}} = (\vec{b}_1 + \vec{b}_2)/2$ which is
(initial) position
of electron/positron in the impact parameter space.
This will be useful when considering motion of electron/positron
in the EM field of nuclei.
The energies of photons are included by the relation: 
$\omega_{1,2}=W_{\gamma\gamma}/2 \exp(\pm Y_{e^+e^-})$.
In the following for brevity we shall use $b_x, b_y$ instead of 
$\overline{b_x}, \overline{b_y}$.
Then $(b_x, b_y)$ is the position in the impact parameter plane,
where the electron and positron are created.{\footnote{Expression (\ref{EPA}) allows to estimate not only 
the lepton pair production but also a production of any other 
particle pair \cite{Klusek-Gawenda:2010vqb}.}}
The differential (in rapidity and transverse momentum) cross section 
could be obtained in each emission point 
in the impact parameter space ($b_x,b_y$).
The transverse momentum distribution of $e^+$ or $e^-$ can be obtained by binning 
in Eq. (\ref{EPA}) $p_t = p^* \sqrt{1-\cos^2 \theta}$ where $p^*$ is lepton momentum in the $e^+e^-$ CM system. Similar binning is done for $y_{e^+}$ and $y_{e^-}$.

The simple $b$-space EPA formula (\ref{EPA}) gives good description 
of dielectron invariant mass distributions \cite{Klusek-Gawenda:2016suk} 
as measured by the ALICE collaboration \cite{ALICE} for  $p_{t,e} >$ 2 GeV. Small electron/positron transverse momenta were not measure so far. 
In this approach the $p_{t,pair}$ and acoplanarity distributions
are Dirac delta functions. In general, this approach allows to describe
rather single lepton distributions (rapidity, transverse momentum).
Below we shall compare a transverse momentum distribution of
electrons of the $b$-space EPA with its counterpart for the
Wigner-function approach \cite{Wigner} for a given range 
of impact parameter, relevant for the discussion in the present paper.

\begin{figure}[!h]
	\includegraphics[scale=0.3]{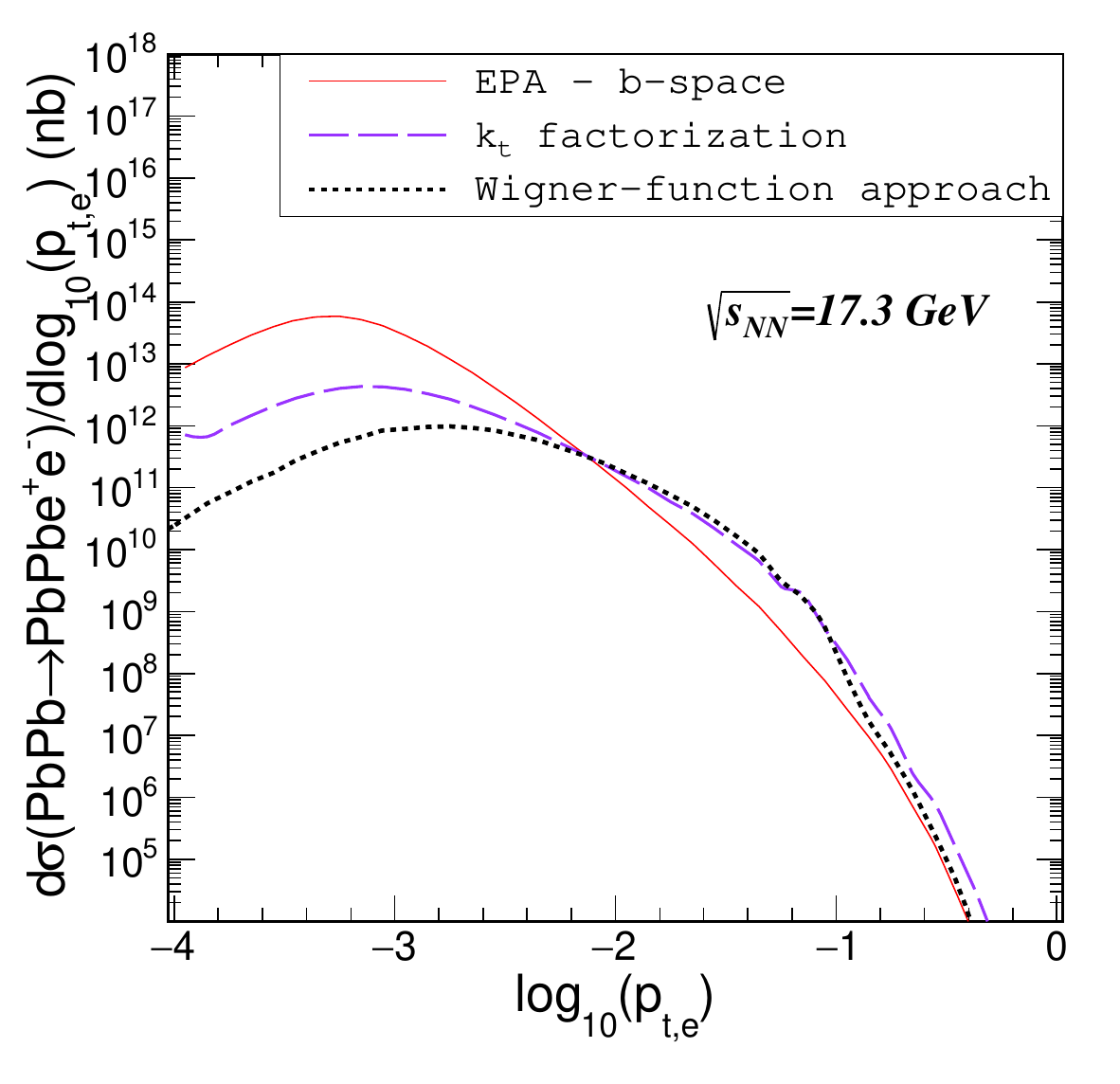}
	\includegraphics[scale=0.3]{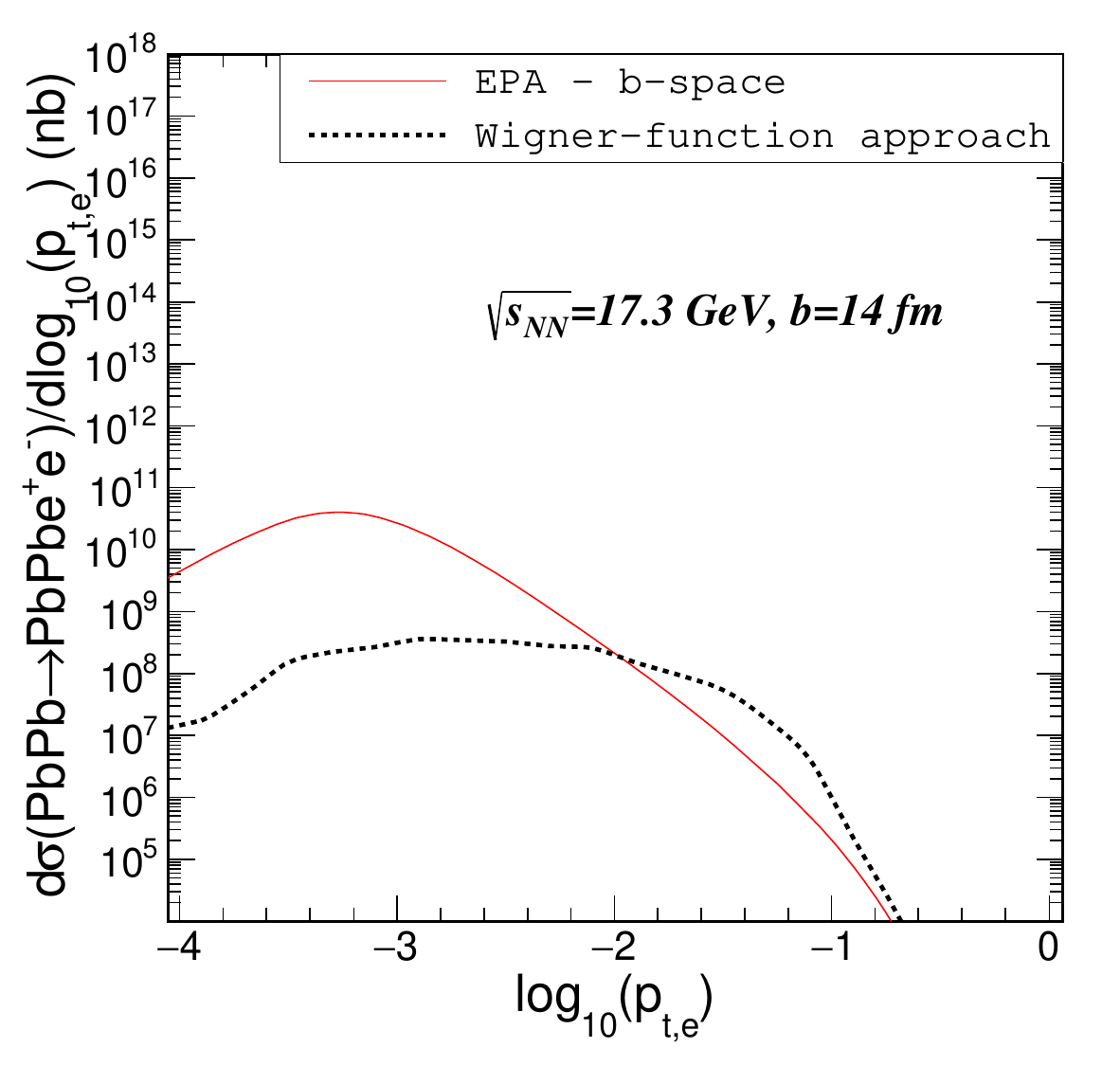}\vspace{-2cm}\\
     \vspace{-2.5cm}\hspace{1.50cm}{\bf(a)}\hspace{6.0cm}{\bf(b)}\\
\vspace{3.25cm}
	\caption{Differential cross section for PbPb$\to$PbPb$e^+e^-$ as a function of $\log_{10}(p_{t,e})$. Shown are results of different approaches as explained in the figure legend. The left panel shows impact parameter integrated cross section while the right panel is for a narrow range of impact parameter $b \in (13.95,14.05)$~fm.}
	\label{fig:dsig_dpt_comparison}
\end{figure}

Can the subtle impact parameter -- lepton transverse momentum
correlations influence the single lepton transverse momenta?
This was not discussed so far in the literature.
In Fig.\ref{fig:dsig_dpt_comparison} we compare results for 
the somewhat simplified $b$-space EPA, the $k_t$- factorization approach \cite{k_t-factorization} and the Wigner-function approach \cite{Wigner}. 
We show results for the full range of impact parameter (left panel) and well as for 
very limited range discussed in the present paper (right panel).
In the latter case there is no $k_t$-factorization result shown in the left panel.
We see interesting results for $p_{t,e}<0.01$~GeV. This region requires dedicated
studies in the context of possible measurement with ALICE3 detector. 
We observe deviations
of the $b$-space EPA result and the Wigner-function approach result
for small lepton transverse momenta $p_t <$ 0.1 GeV.
Sizeable numerical fluctuations in the Wigner-function approach
can be observed.
Since in the present exploratory calculations we are interested
rather in estimating the size of the outcome related to the electromagnetic final state interactions (FSI) of leptons with the nuclei as well as phase-space
localization (rapidity, transverse momentum) of the new FSI effect, 
in the present paper 
we wish to use a more handy $b$-space EPA approach 
(6-dimensional integration). One has to have 
in mind that precise evaluation of the low-$p_t$ distributions requires 
use of the rather complicated Wigner-function approach 
(10-dimensional integration).
Using the Wigner-function approach together with the inclusion
of the final state interaction effects goes beyond the scope
of the present paper but would not change the general conclusions
drawn here.

%
\begin{figure}[!hbt]
     \includegraphics[scale=0.5]{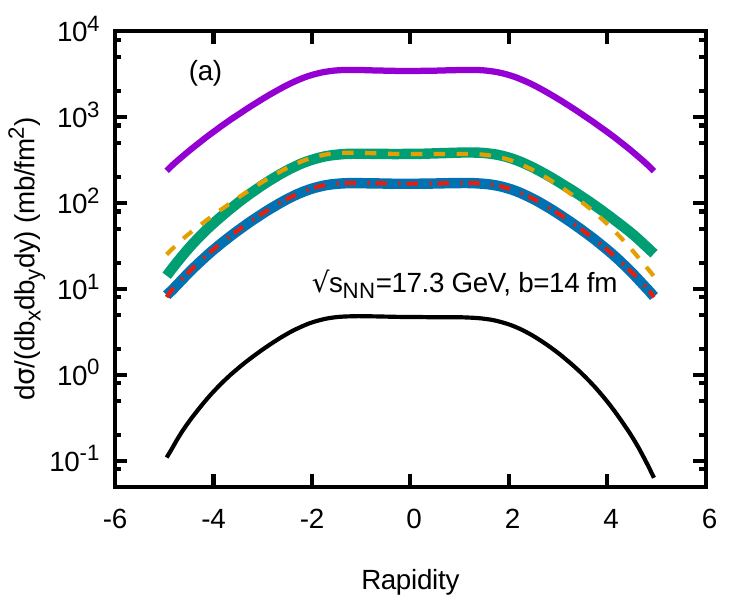}
     \includegraphics[scale=0.5]{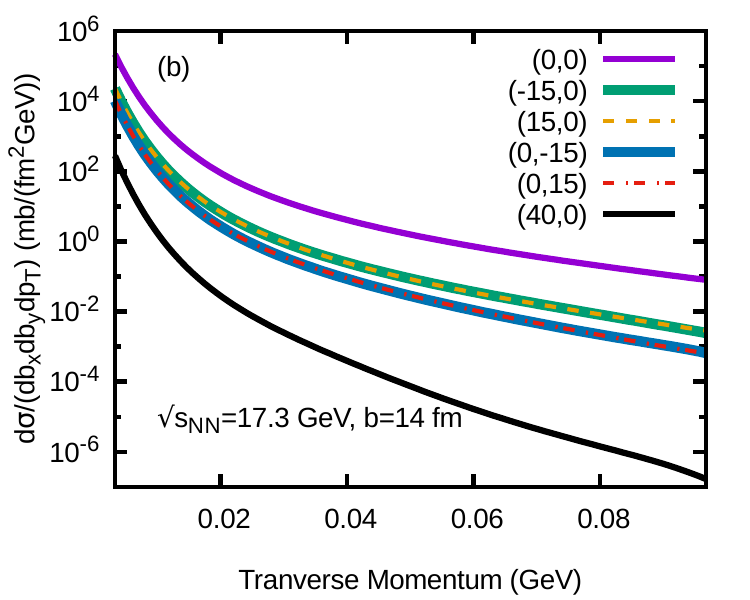}
     \caption{(Color on-line) The differential cross section for various emission points
       of electrons/positrons produced in the $^{208}$Pb+$^{208}$Pb
       reaction at 158~GeV/nucleon energy ($\sqrt{{s}_{NN}} =$ 17.3~GeV)
       at impact parameter 14$\pm$0.05~fm. The cross section 
       for selected points ($b_x,b_y$): (0, 0), ($\pm$15~fm, 0), (0, $\pm$15~fm) and (40~fm, 0)
       are integrated over $p_T$ (a) and rapidity (b), respectively.}
     \label{fig02b}
\end{figure}

The calculations will be done assuming the collision of
$^{208}$Pb+$^{208}$Pb 
at 158~GeV/nucleon energy ($\sqrt{{s}_{NN}} =$ 17.3~GeV) corresponding to
the CERN SPS and $\sqrt{{s}_{NN}} =$ 200~GeV of 
the STAR RHIC at impact parameter 14$\pm$0.05~fm which is 
approximately twice the radius of the lead nucleus. 
This is minimal configuration assuring ultra-peripheral collisions.

Figure \ref{fig02a} illustrates the differential cross section on 
the plane of rapidity ($y$) vs. transverse momentum ($p_T$). 
Rather broad range of rapidity (-5, 5) is chosen, but the distribution in $p_T$ will be limited
to (0, 0.1~GeV) as the cross section drops at $p_T$ = 0.1~GeV 
already a few orders of magnitude. The electromagnetic effects may 
be substantial only in the region of the small transverse momenta.

Thus for our exploratory study here we have limited the range for rapidity 
to (-5, 5) and for transverse momentum to $p_T$=(0, 0.1~GeV).
The integrated distribution can be seen in Fig.~\ref{fig02b}(a) and
(b). There we compare the distributions obtained for 
different emission points ($b_x,b_y$): (0, 0), ($\pm$15~fm, 0), 
(0, $\pm$15~fm) as shown in Fig.~\ref{fig01}.
The behavior of the differential cross section is very similar in each 
($b_x,b_y$) point but it differs in normalization as it is shown 
in Fig.~\ref{fig02b}.

\begin{figure}[!hbt]
\resizebox{0.4\textwidth}{!}{%
\includegraphics{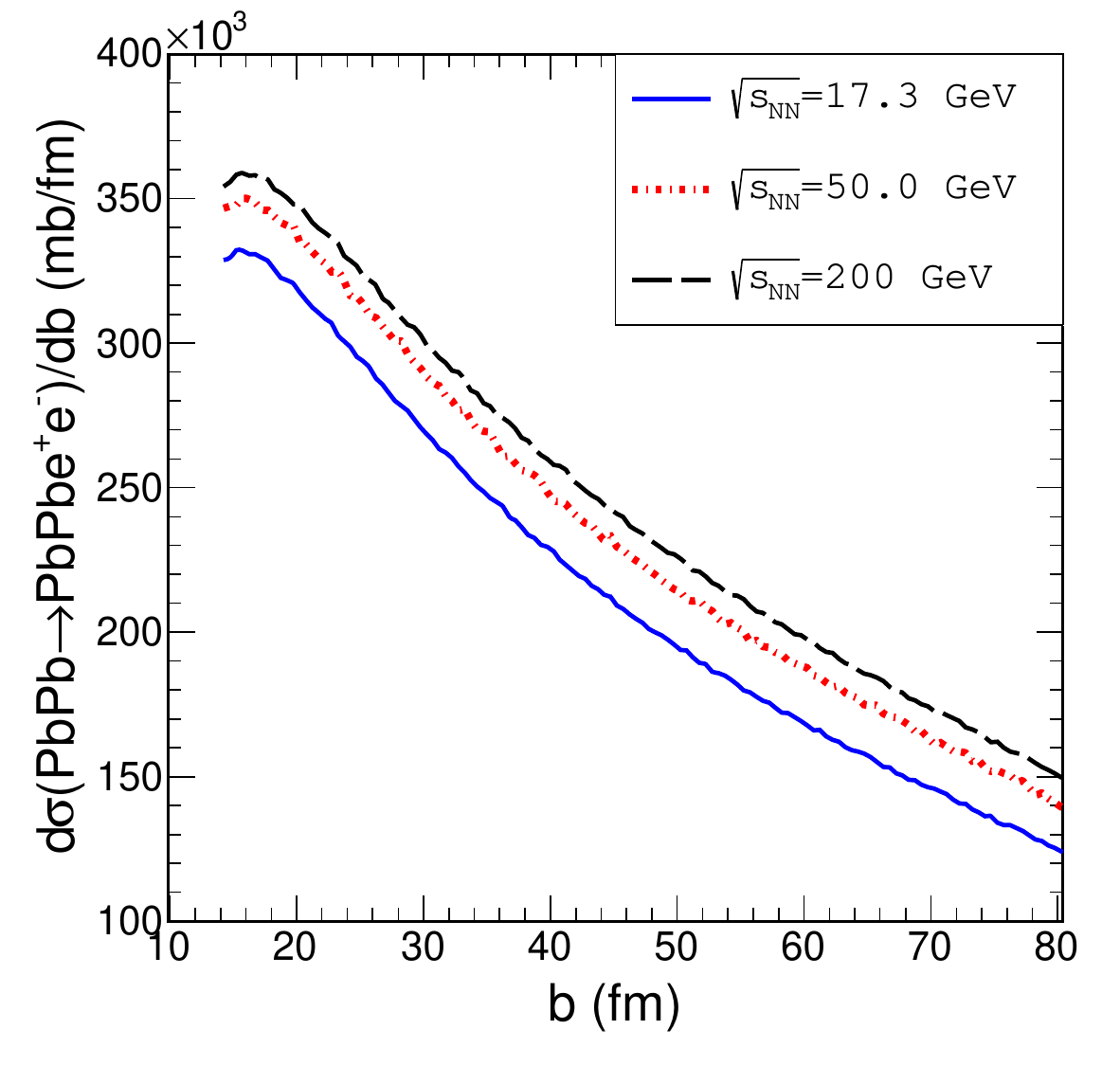}}
\caption{
Distribution of the cross section in the impact parameter $b$
for different energies:
$\sqrt{s_{NN}}$ = 17.3, 50 and 200~GeV (from bottom to top).
}
\label{fig:dsig_dbm}
\end{figure}

In Fig.~\ref{fig:dsig_dbm} we show a distribution of the cross section
in impact parameter $b$ for different collision energies
$\sqrt{s_{NN}}$ = 17.3, 50, 200~GeV.
In this calculation we have taken $p_T >$ 0~GeV (the cross section
strongly depends on the lowest value of lepton transverse momentum $p_T$).
In general, the larger collision energy the broader the range of impact
parameter. However, the cross section for $b \approx R_{A_1} + R_{A_2}$
is almost the same. Only taking into account limitation, e.g. on the momentum transfer, makes the difference in the cross section significant even at $b=14$~fm.

\begin{figure}[!hbt]
     \includegraphics[scale=0.3]{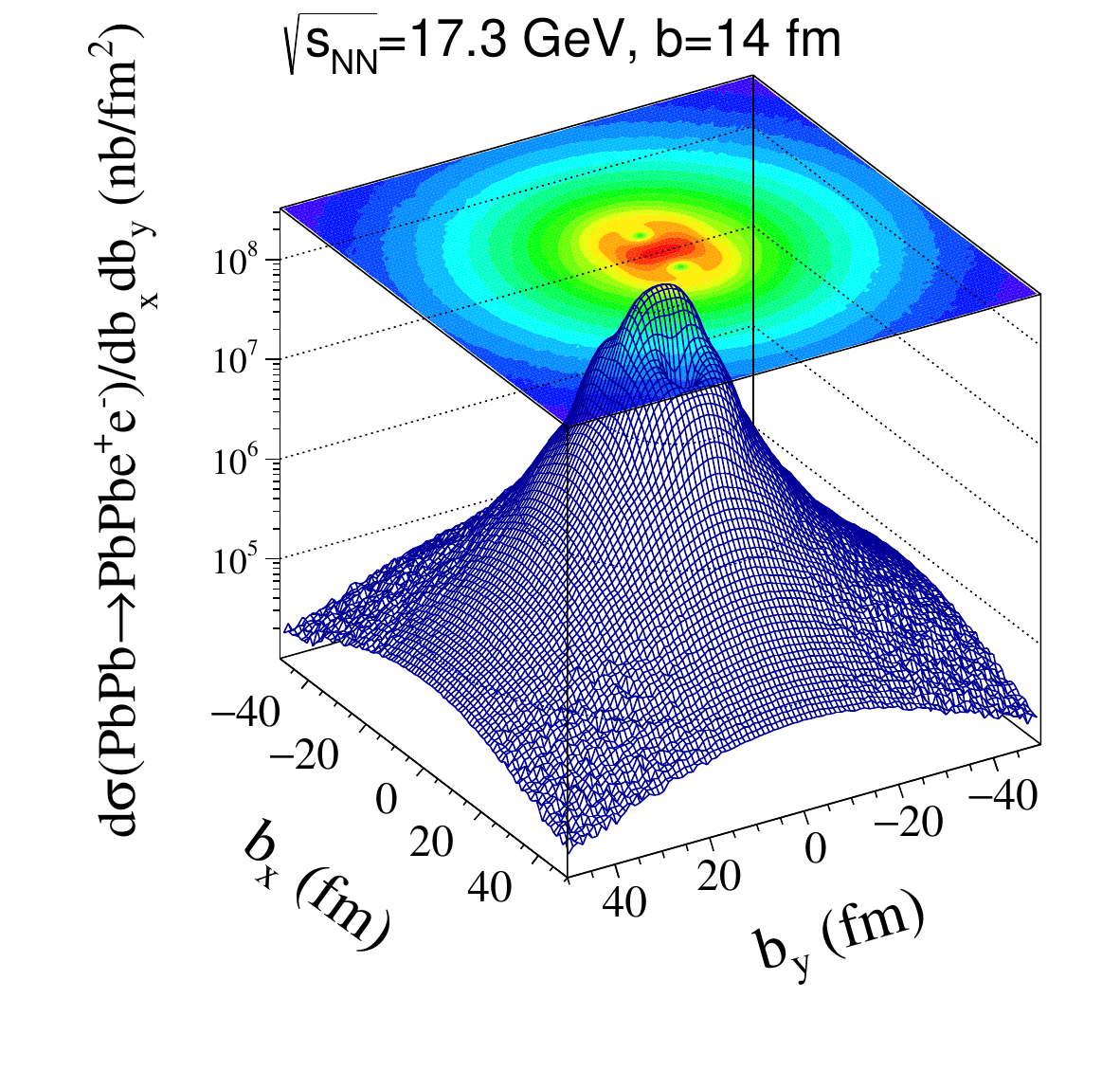}
     \includegraphics[scale=0.3]{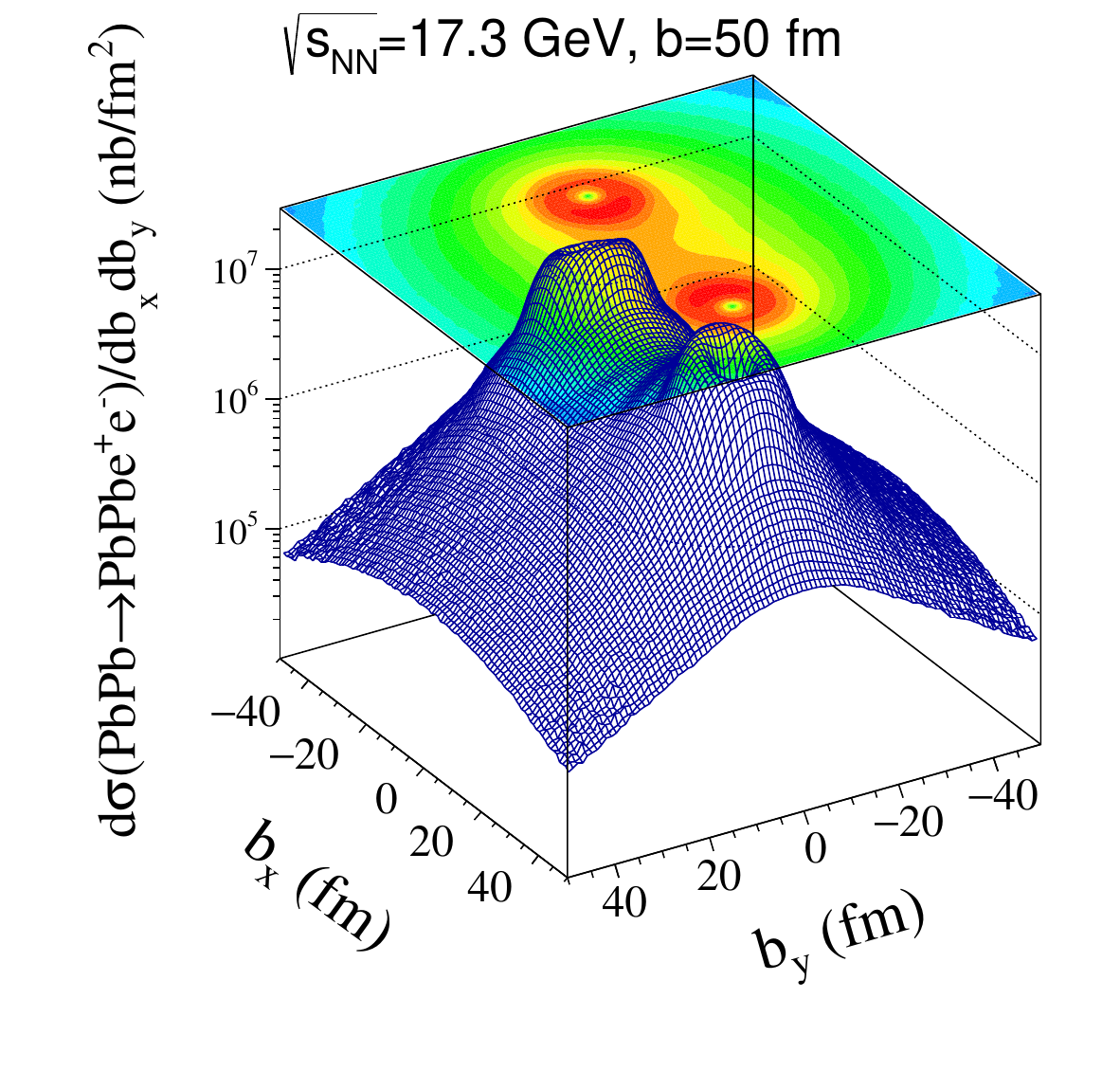}\vspace{-2cm}\\
     \vspace{-6cm}\hspace{4.0cm}{\bf(a)}     \hspace{5.0cm}{\bf(b)}\\
     \vspace{7cm}
     \caption{ Two-dimensional cross section
      as a function of $b_x$ and $b_y$ for
      two values of impact parameter: 
      (a) b=14$\pm$0.05~fm and (b) b=50$\pm$0.05~fm.}
     \label{fig02d}
\end{figure}

The emission point of the electrons/positrons does not change the
behavior (shape) of the cross section on the ($y,p_T$) plane but 
changes the absolute value of the cross section.
As it is visible in Fig.~\ref{fig02b} (a) and (b) the biggest
cross section is obtained for the CM emission point. The production of 
$e^+,e^-$ at ($b_x,b_y$) = (40~fm, 0) i.e. far from the CM point, 
is hindered by three orders of magnitude. 
Moreover, the production at $b_x$=$\pm$15~fm and $b_y$=0 is more
preferable than the production at $b_x$=0 and $b_y$=$\pm$15~fm 
what is fully understandable taking into account the geometry of 
the system (see Fig.~\ref{fig01}). 
As the system taken here into consideration is fully symmetric
($A_1 = A_2, Z_1 = Z_2$), 
thus corresponding results are symmetric under the following
replacements: $b_x \to -b_x$ or $b_y \to -b_y$.

Figure~\ref{fig02d} compares the integrated cross section on reaction 
plane ($b_x,b_y$) for two impact parameters: (a) b=14$\pm$0.05~fm 
(when nuclei are close to each other) and (b) b=50$\pm$0.05~fm (when nuclei are well separated). 
The landscape reflects the position of the nuclei in the moment of
the closest approach.
Similar plots have also been done for higher $\sqrt{s_{NN}}$ but 
the shape is almost unchanged, only the cross section value is
different. 
This figure illustrates the influence of the geometry of the reaction. 
Regardless of the impact parameter (b), distance between colliding nuclei, the cross-section has
a maximum at $b_x=0$. The change of $b$ is correlated with the shift 
in a peak at $b_x$.

%
\begin{figure}[!hbt]
\resizebox{0.90\textwidth}{!}{%
     \includegraphics{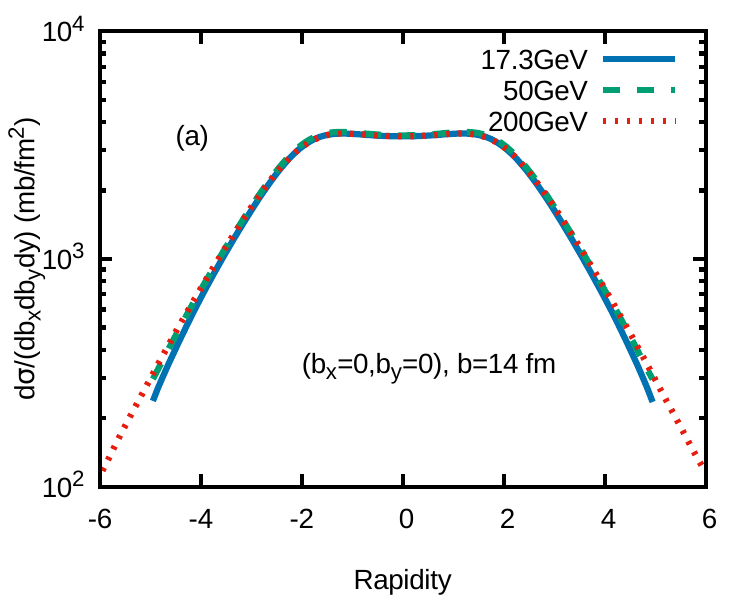}
     \includegraphics{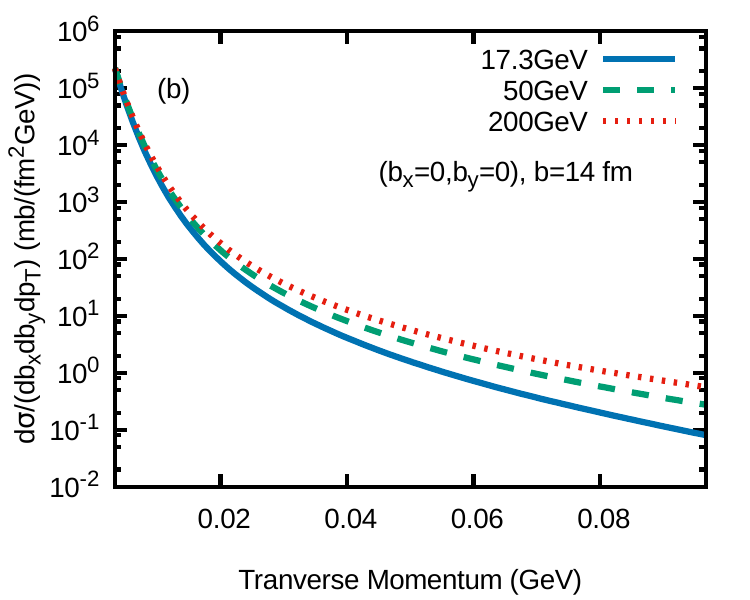}}
     \caption{(Color on-line) The differential cross section for various emission points
       of electrons in the $^{208}$Pb+$^{208}$Pb reaction 
       at $\sqrt{{s}_{NN}} =$ 17.3, 50, 200~GeV at impact parameter 14$\pm$0.05~fm.}
     \label{fig02c}
 \end{figure}
%

The calculations confirm that the shape of the electron/positron 
distribution, shown in Fig.~\ref{fig02c} does not depend on 
the energy of the colliding nuclei. There are visible small differences 
in the magnitude of cross sections 
for $\sqrt{{s}_{NN}} =$ 17.3 and 200~GeV
at least in the selected limited $p_T$=(0, 0.1)~GeV range. 
Dependence on the rapidity is even weaker as the differences are visible
only for $|y|>$3.

These cross sections are used as weights in calculation of electromagnetic
effects between electrons/positrons and the fast moving nuclei. The
corresponding matrix has following dimensions: $b_{x,y}$=(-50~fm, 50~fm) - 
99$\times$99 points in the reaction plane and 100$\times$15 in the ($y,p_T$) space.

\begin{figure}[!hbt]
\resizebox{0.49\textwidth}{!}{%
		\includegraphics{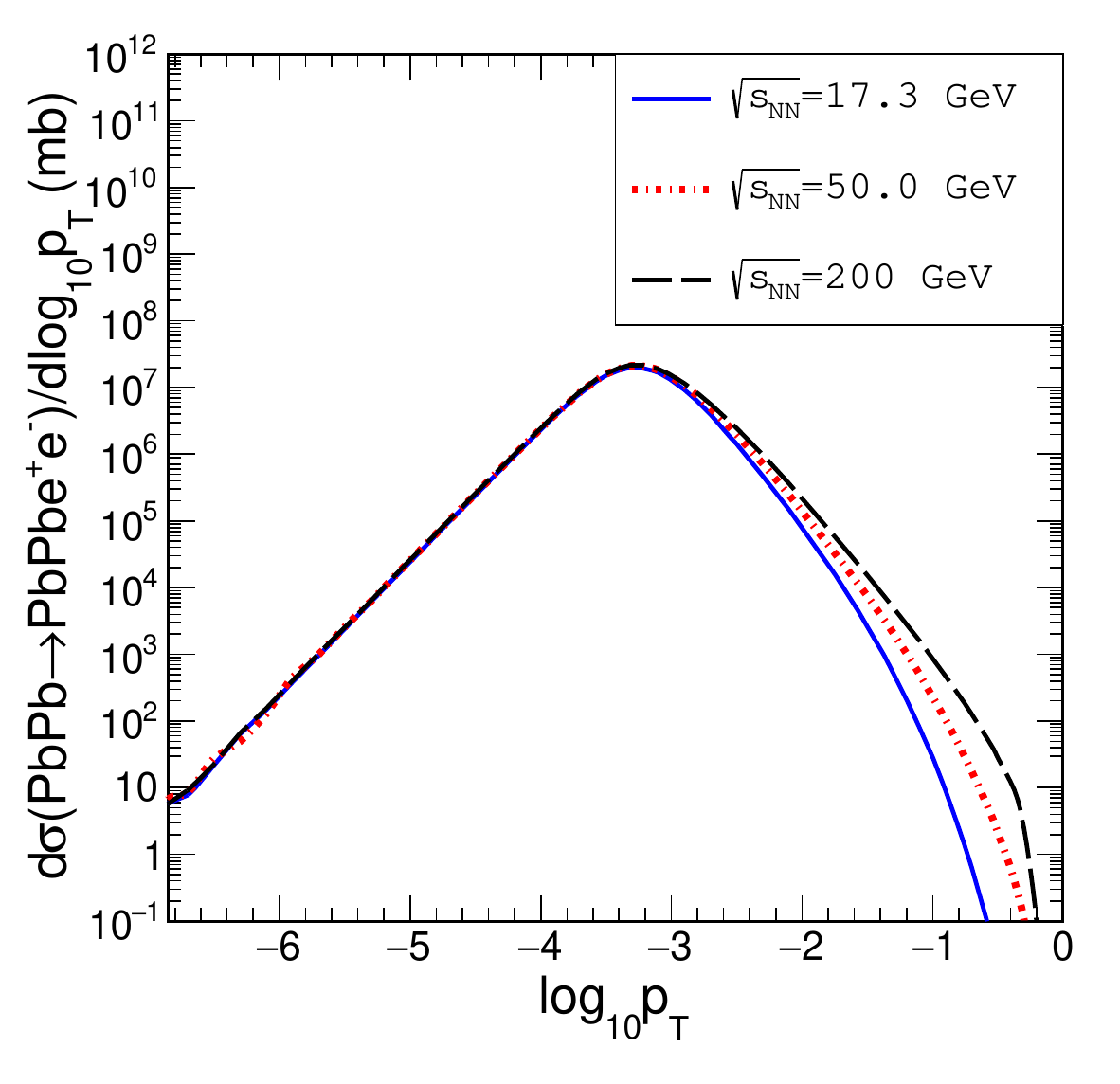}}
		\caption{
			Distribution of the cross section in $\log_{10}p_T$
			for different energies:
			$\sqrt{s_{NN}}$ = 17.3, 50 and 200~GeV (from bottom to top).
		}
		\label{fig:dsig_dxipt}
\end{figure}

Rather small transverse momenta enter such calculation.
To illustrate this in Fig.~\ref{fig:dsig_dxipt} we show a distribution
in $log_{10}(p_T)$. As seen from the figure the cross section is
integrable and we have no problem with this with our Monte Carlo
routine \cite{Lepage:1977sw}.

\section{Electromagnetic interaction effects}  \label{Sect.III}

It is our aim to study the influence of the strong EM fields on
the distributions of produced electrons or positrons, separately for
electrons and positrons.

The spectator system are modeled as two uniform spheres in 
their respective rest frames that change into disks in the 
overall center-of-mass collision frame.
\begin{figure}[!hbt]
\resizebox{0.99\textwidth}{!}{%
\includegraphics{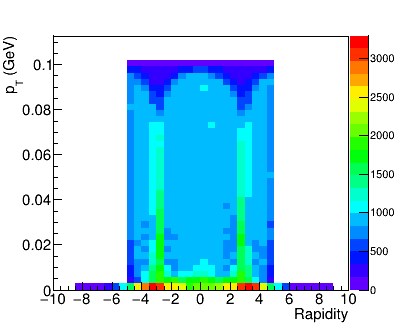}
\includegraphics{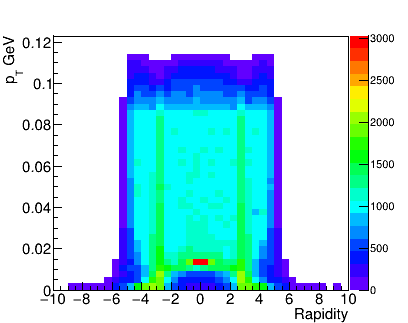}
\includegraphics{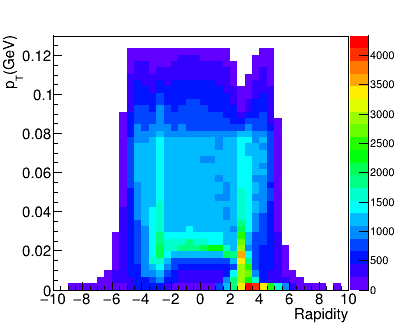}}\\
\vspace{-5.3cm}\hspace{0.cm} {\bf (a) ($b_x$=0, $b_y$=0) \hspace{0.5cm} (b) ($b_x$=0, $b_y$=15~fm) \hspace{0.5cm} (c) ($b_x$=15~fm, $b_y$=0)} \vspace{5.55cm}\\
\resizebox{0.99\textwidth}{!}{%
\includegraphics{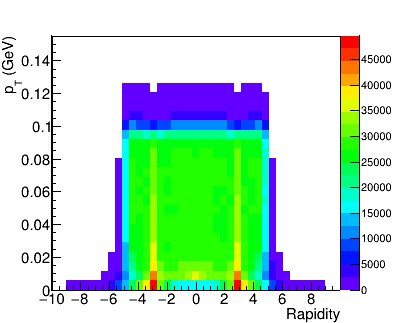}
\includegraphics{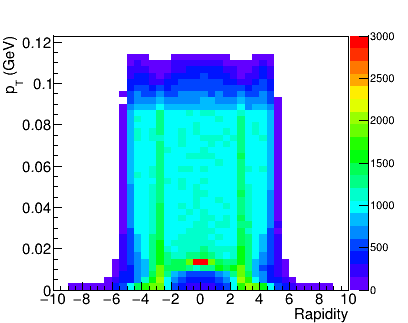}
\includegraphics{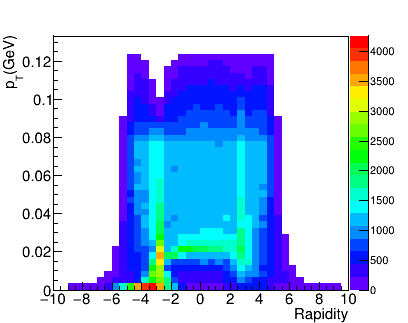}}\\
\vspace{0.3cm}\hspace{0.cm} {\bf (d) full ($b_x$, $b_y$)  \hspace{0.5cm} (e) ($b_x$=0, $b_y$=-15~fm) \hspace{0.5cm} (f) ($b_x$=-15~fm, $b_y$=0)} \vspace{-0.cm}
\caption{(Color on-line) Rapidity vs $p_T$ distributions for final (subjected to EM
  effects) electrons for different emission points: 
(a) (0, 0) and (d) full xy plane; (b)
(0, 15~fm); (c) (15~fm, 0); (e) (0, -15~fm) and (f) (-15~fm, 0).
These results are for $\sqrt{s_{NN}}$ = 17.3~GeV.}\label{fig03}
\end{figure}
The total charge of nuclei is 82 consistent with UPC. 
The lepton emission region is reduced to a single point and the time of emission is a free parameter. 

\begin{figure}
\resizebox{0.99\textwidth}{!}{%
\includegraphics{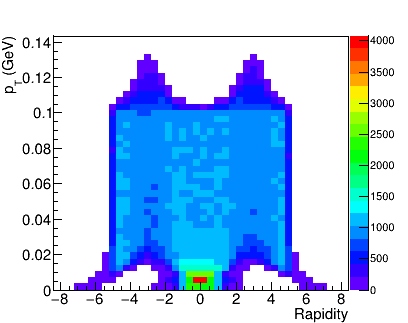}
\includegraphics{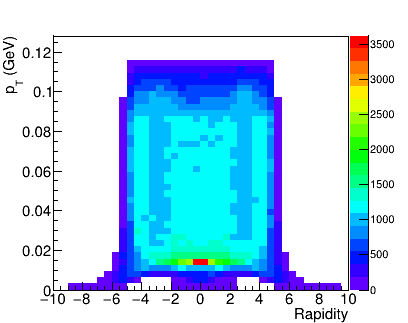}
\includegraphics{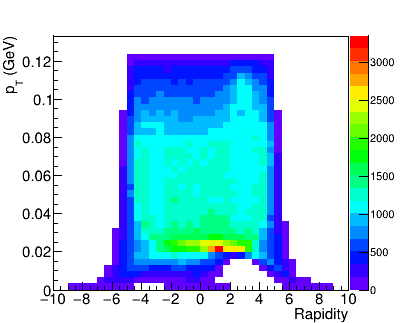}}\\
\vspace{-0.3cm}\hspace{0.cm} {\bf (a) ($b_x$=0, $b_y$=0) \hspace{0.5cm} (b) ($b_x$=0,$b_y$=15~fm) \hspace{0.5cm} (c) ($b_x$=15~fm, $b_y$=0)}  \vspace{0.55cm}\\
\resizebox{0.99\textwidth}{!}{%
\includegraphics{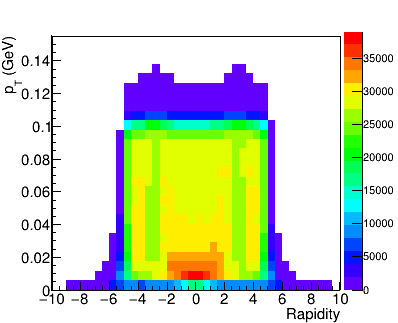}
\includegraphics{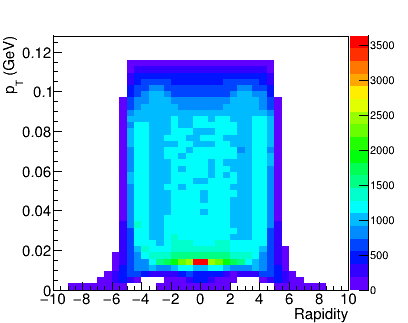}
\includegraphics{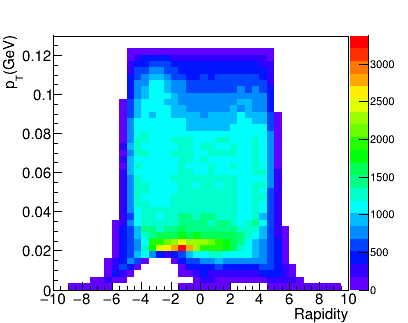}}\\
\vspace{-0.3cm}\hspace{0.6cm} {\bf (d) full ($b_x$, $b_y$) \hspace{0.5cm} (e) ($b_x$=0, $b_y$=-15~fm) \hspace{0.5cm} (f) ($b_x$=-15~fm, $b_y$=0)} \hspace{0cm}\\
\vspace{0.cm}
\caption{(Color on-line) Rapidity vs $p_T$ distributions for final (subjected to EM 
effects) positrons for different emission points: 
(a) (0, 0) and (d) full xy plane; (b) (0, 15~fm); (c) (15~fm, 0);
(e) (0, -15~fm) and (f) (-15~fm, 0).
These results are for $\sqrt{s_{NN}}$ = 17.3~GeV.}\label{fig04}
\end{figure}

\begin{figure*}[!hbt]
\resizebox{0.99\textwidth}{!}{%
\includegraphics{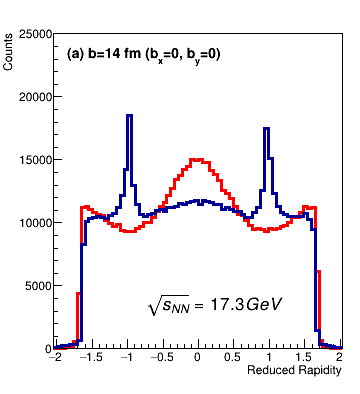}
\hspace{0.cm}
\includegraphics{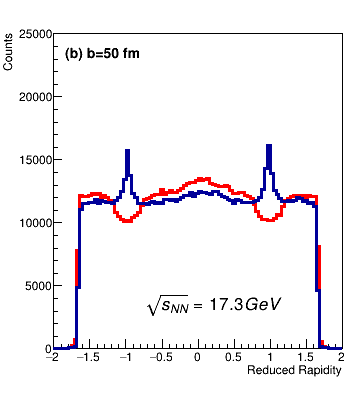}}\vspace{-0cm}\\
\vspace{0cm}
\resizebox{0.99\textwidth}{!}{%
\includegraphics{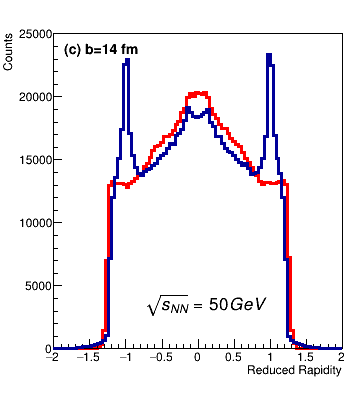}
\includegraphics{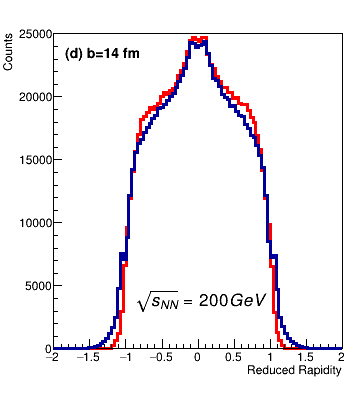}}\vspace{-0cm}
\vspace{0cm}
\caption{(Color on-line) Reduced rapidity distributions for final electrons (blue) and
  positrons (red) for fixed $b$ and ($b_x$=0,$b_y$=0) plane of
emission points at three collision energies: $\sqrt{s_{NN}}$=17.3, 50 and 200~GeV.
}
     \label{fig05a}
\end{figure*}

In this work we assume there is no delay time between collisions of
nuclei and the start of the EM interactions. 
The $z$-dependence of the first occurrence of the $e^+ e^-$ pair is
beyond the EPA and is currently not known.
In our opinion production of $e^+ e^-$ happens when the moving
cones, fronts of the EM fields, cross each other. This happens
for $z \approx$ 0. 
In the following we assume $z$ = 0 for simplicity.
\footnote{Any other distribution could be taken.}

%
\begin{figure*}[!hbt]
\resizebox{0.99\textwidth}{!}{%
\includegraphics{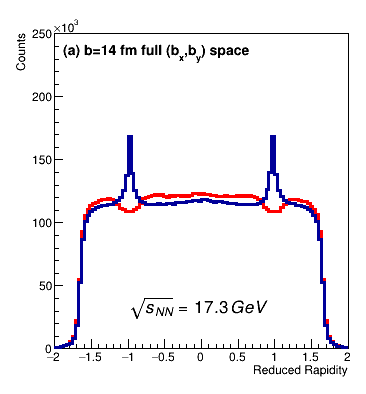}
\includegraphics{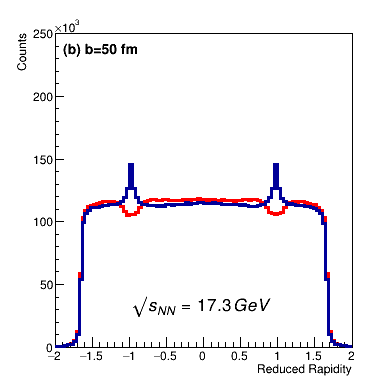}}\\
\resizebox{0.99\textwidth}{!}{%
\includegraphics{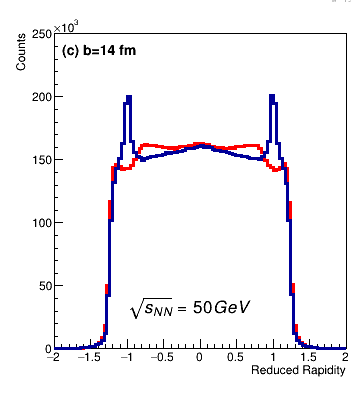}
\includegraphics{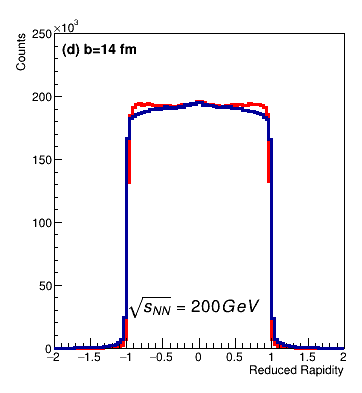}}
\caption{(Color on-line) Reduced rapidity distributions for final electrons (blue) 
and positrons (red) for fixed $b$ and full ($b_x$,$b_y$) plane of
emission points at three collision energies: $\sqrt{s_{NN}}$=17.3, 50 and 200~GeV.}
     \label{fig05b}
\end{figure*}

The trajectories of $e^{\pm}$ in the field of moving nuclei are obtained
by solving the equation of motion numerically for electrons/positrons:
\begin{equation}
\frac{d \vec{p}_{e^{\pm}}}{d t} =
\vec{F}_{1,e^{\pm}}(\vec{r}_1,t) + \vec{F}_{2,e^{\pm}}(\vec{r}_2,t) \; .
\label{equation_of_motion}
\end{equation}

The total interaction is a superposition of interactions with both
nuclei which positions depend on time.
We solve the motion of electron/positron in the overall center 
of mass system, i.e. both position and time are given in this frame.
In this frame we have to deal with both electric and magnetic force
\cite{Rybicki:2006qm}.
Because nuclei are very heavy compared to electrons/positron
their motion is completely independent and is practically not distorted
by the EM interaction.
We take:
\begin{eqnarray}
\vec{r}_1(t) =  + {\hat z} c t + \vec{b}/2 \; , \nonumber \\
\vec{r}_2(t) =  - {\hat z} c t - \vec{b}/2 \; ,
\end{eqnarray} 
i.e. assume that the nuclei move along straight trajectories
independent of the motion of electron/positron.
The step of integration depends on energy and must be carefully adjusted.

Our approach can also be used to study the influence on the observables 
related to the dilepton pair, such as invariant mass or transverse	
momentum of the pair. Temporary calculation shows that the effect
then is much smaller than for individual distributions.
Therefore we decided not to discuss this in this paper.

The rapidity vs $p_T$ distributions of initial leptons are 
obtained by randomly choosing the position on the two-dimensional space. 
The path of particle in electromagnetic field generated by nuclei
are traced up to 10 000~fm away from the original interaction point. 
The Monte Carlo method is used to randomize the 
initial rapidity and $p_T$ from uniform distribution. 
The initial rapidity and $p_T$ of electrons/positrons are randomly
chosen in the range: $y$=(-5,5) and $p_T$=(0, 0.1)~GeV as fixed
in the previous section. The ($y,p_T$) distributions for final
(subjected to the EM evolution) leptons
are presented in Fig.~\ref{fig03} for electrons and in Fig.~\ref{fig04} 
for positrons. These distributions were obtained by analyzing 
the EM evolution event-by-event.
The number of events taken here is $n_{event}$ = 10$^7$ 
for each two-dimensional plot.

For electron production an enhancement and for positron a loss with
respect to the neighborhood or/and flat initial distribution
is observed for $y \approx \pm$ 3. This corresponds to the beam rapidity
at $\sqrt{s_{NN}}$= 17.3~GeV  energy.

These two sets of two-dimensional plots illustrate 
effect of the EM interaction between $e^+$ or $e^-$ and the moving nuclei.  
The motion of particles in the EM field of nuclei changes 
the initial conditions and the final ($y,p_T$) are slightly different.
Fig.~\ref{fig03} shows the behavior of electrons and 
Fig.~\ref{fig04} of positrons at CM (a) and 
different impact parameter points: (panels c,f) ($\pm$15~fm,0) and 
(panels b,e) (0,$\pm$15~fm) marked in Fig.~\ref{fig01}. 
We observe that the maximal number of electrons is located 
where the cross section for positrons has minimum. The Coulomb influence is well visible as 
a missing areas for positrons for $p_T<0.02$~GeV for particles
emitted from the CM point. The emission of leptons from $b_y$=$\pm$15~fm gives
lower effect. The asymmetry in emission is well visible for
$b_x$=$\pm$15~fm, 
where a larger empty space is for positive rapidity when $b_x$=15~fm and 
for negative rapidity when $b_x$=-15~fm. 

Although the EM effects are noticeable for 
electrons/positrons in different impact parameter points, 
the integration over full reaction plane washes out almost totally this impact.

For comparison the results of integration over full space 
$b_x$=(-50~fm, 50~fm) and $b_y$=(-50~fm, 50~fm) are shown 
in panels (d) of Figs.~\ref{fig03} and \ref{fig04}. 
These results are independent of the source of leptons, thus it could 
be treated as a general trend and an indication for which 
rapidity-transverse momentum ranges one could observe effects of the EM
interaction between leptons and nuclei.
%
\begin{figure*}
\resizebox{0.99\textwidth}{!}{%
\includegraphics{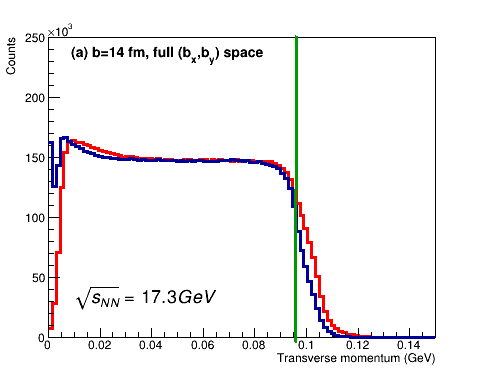}\hspace{-0.cm}
\includegraphics{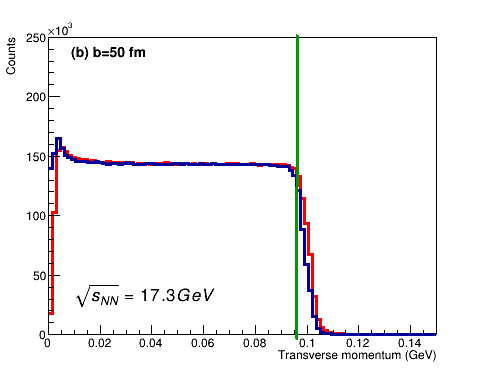}}\\
\resizebox{0.99\textwidth}{!}{%
\includegraphics{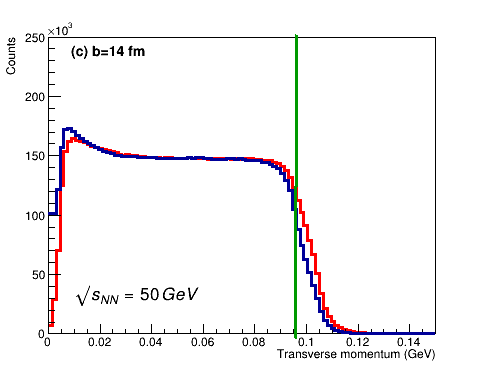}\hspace{-0.cm}
\includegraphics{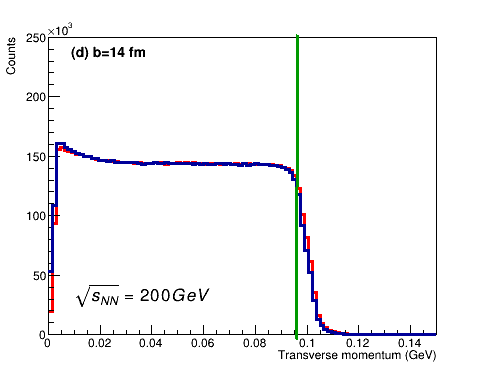}}
\caption{(Color on-line) Transverse momentum distributions for final electrons (blue) 
and positrons (red) for fixed $b$ and full ($b_x$, $b_y$) plane of
emission points at three collision energies: $\sqrt{s_{NN}}$=17.3, 50 and 200~GeV.}
     \label{fig05c}
\end{figure*}

%
The distribution in reduced rapidity of final leptons are shown 
in Fig.~\ref{fig05a} and \ref{fig05b}. The reduced rapidity
(dimensionless quantity) is the rapidity $y$ normalized to the beam rapidity $y_{beam}$
(different for various collision energies)
\begin{equation}
y_{red} = y / y_{beam} \; ,
\end{equation}
where
\begin{equation}
y_{beam} = \pm ln \left( \frac{\sqrt{s_{NN}}}{m_p} \right)
\end{equation}
and $m_p$ is the proton mass.
The results shown above were obtained, somewhat arbitrarily, with
uniform distribution in $(y,p_T)$.
This leads to the observation of peaks or dips at beam rapidities.
No such peaks appear for $\sqrt{s}$ = 200~GeV as here the chosen range
of rapidity (-5,5) is not sufficient.
Whether such effects survive when weighting with the b-space EPA
cross section will be discussed below.

Fig.~\ref{fig05a} is focused on emission from the center of mass point 
and Fig.~\ref{fig05b} is obtained when integrating over full ($b_x,b_y$) plane. 
The main differences between electron (blue lines) and 
positron (red lines) distributions are not only at midrapidities 
but also around the beam rapidity. The effect is more visible 
for the CM emission point but it is slightly smoothed out when 
the full $(b_x, b_y)$ plane is taken into consideration.   
Moreover increasing the impact parameter (panels (a) and (b)) 
diminishes the difference between rapidity distributions of final 
electrons and positrons. The beam energy is another crucial parameter. 
The collision with $\sqrt{s_{NN}} > 100$~GeV (panel (d)) does not 
allow for sizeable effects of electromagnetic interaction between 
leptons and nuclei, at least at midrapidities.

The discussion of the EM interaction between $e^+e^-$ and nuclei has 
to be completed by combining with the cross section of lepton production
as obtained within EPA.
Taking into account the leptons coming from photon-photon fusion 
the distributions from Fig.~\ref{fig03} and \ref{fig04} are multiplied 
by differential cross section obtained with Eq.(~\ref{EPA}).
 
The details of the method are presented in Ref.~\cite{Rybicki:2006qm} 
and adapted here from pion emission to electron/positron emission.
\begin{figure*}[!hbt]
\resizebox{0.99\textwidth}{!}{%
     \includegraphics{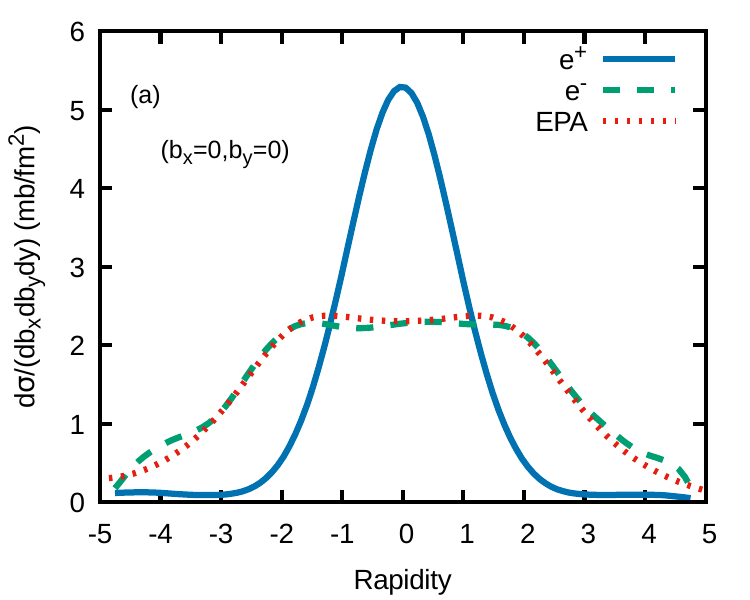}
     \includegraphics{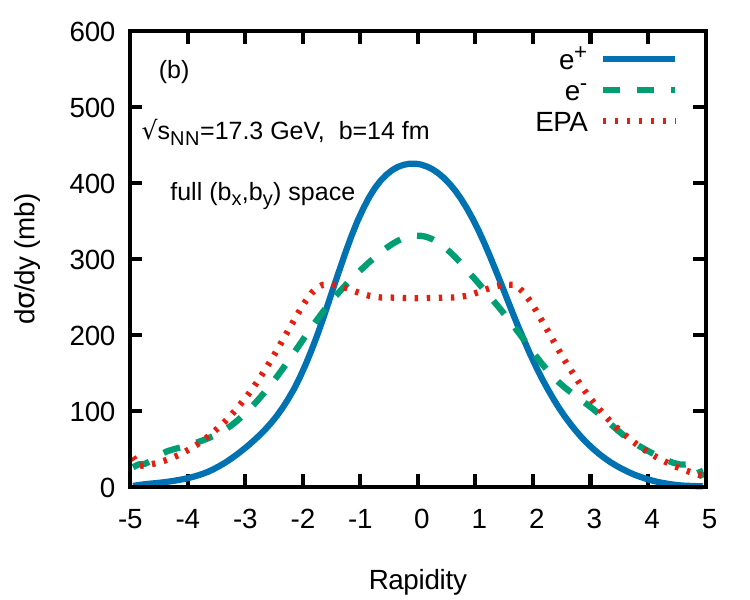}}
     \caption{(Color on-line)
     The electron and positron emission cross section normalized to 100\% 
     in the $^{208}$Pb+$^{208}$Pb reaction at 158~GeV/nucleon energy 
($\sqrt{{s}_{NN}} =$ 17.3~GeV) at impact parameter 14$\pm$0.05~fm
assuming the $p_T$=(0, 0.1)~GeV produced in the center of mass (0, 0)
point (a) and when integrating over full reaction space (b).
Shown are original EPA distributions (dotted line) and results when
including evolution in the EM field of nuclei for positrons (solid line)
and for electrons (dashed line).
}
     \label{fig07}
\end{figure*}
In Fig.~\ref{fig05c} we show the influence of EM interaction on $p_T$ distributions.
Here we integrate over rapidity and ($b_x,b_y$). One can observe that the EM effects lead to a diffusion of transverse momenta 
(see the diffused edge at $p_T=0.1$~GeV, marked by green vertical line). No spectacular influence is observed when changing the 
impact parameter or beam energy.

Fig.~\ref{fig07} shows a comparison of rapidity distribution of 
final electrons and positrons, assuming the particles are emitted 
from (a) the center of mass (0, 0) point and (b) when integrating over 
$b_x,b_y$. 
The comparison is done between EPA distribution relevant for initial stage (red, dotted line) with the final stage, resulting from the EM interaction of charged leptons with positively charged nuclei. 
If leptons are produced in the CM point, the electron
distributions are almost unchanged but positron distributions 
are squeezed to $|y| < 2$. 
If the cross section is integrated over full ($b_x,b_y$) parameter space, 
the positron distribution is still steeper than that for electrons 
but mainly for $|y| <$ 2. 

Even when the leptons produced in the full ($b_x,b_y$) plane are considered, the
$e^+$ and $e^-$ distributions are different from the initial ones. The electrons under the EM interactions are focused at midrapidities.

%
\begin{figure}[!hbt]
\resizebox{0.99\textwidth}{!}{%
\includegraphics{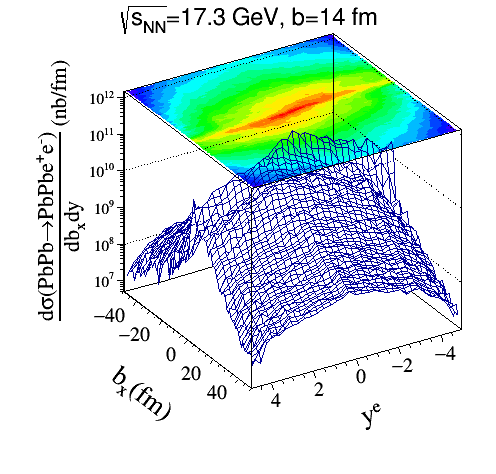}
\includegraphics{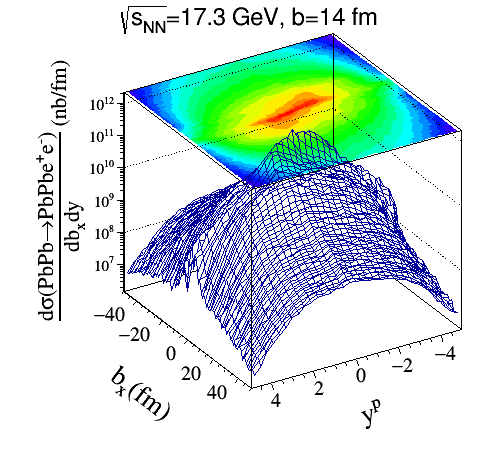}}\vspace{-1.0cm}\\
{\bf\hspace{1.5cm} (a) \hspace{5.0cm} (b)}\\
\vspace{0.5cm}
\resizebox{0.99\textwidth}{!}{%
\includegraphics{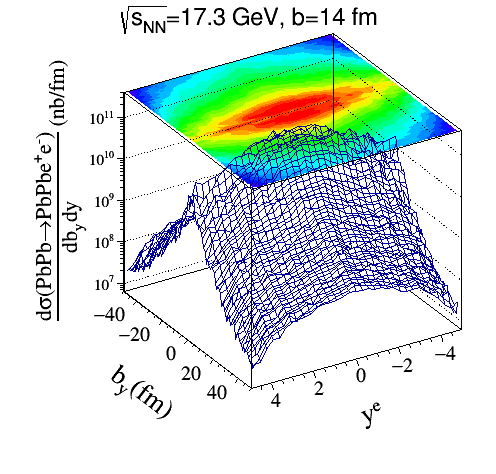}
\includegraphics{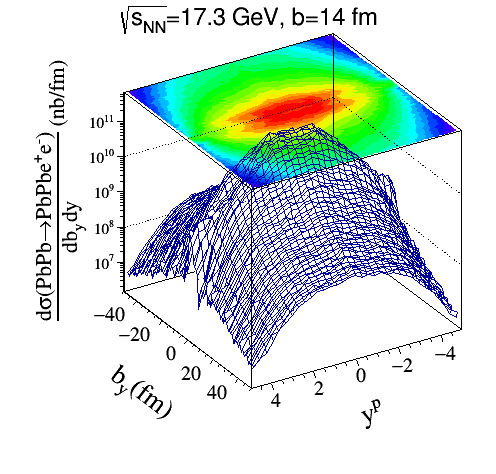}}\vspace{-1.0cm}\\
{\bf\hspace{1.5cm} (c) \hspace{5.0cm} (d)}\\
\vspace{0.5cm}
\caption{ Distribution of electrons ((a), (c)) and positrons ((b), (d)) for $\sqrt{s_{NN}}$= 17.3
 ~GeV at b=14~fm
  integrated over ($b_x,b_y$)=(-50~fm, 50~fm),
  $p_T^{ini}$=(0, 0.1~GeV)}
     \label{fig08}
\end{figure}
\begin{figure}
\resizebox{0.99\textwidth}{!}{%
\includegraphics[scale=0.4]{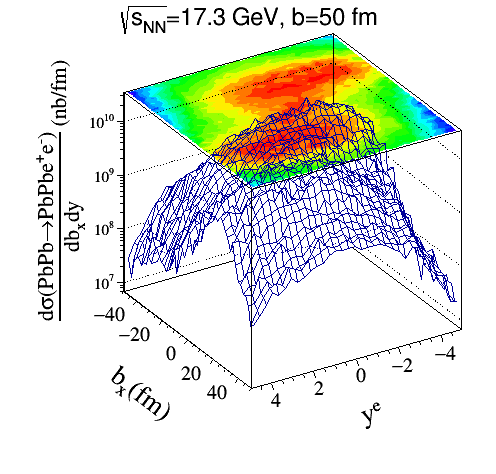}
\includegraphics[scale=0.4]{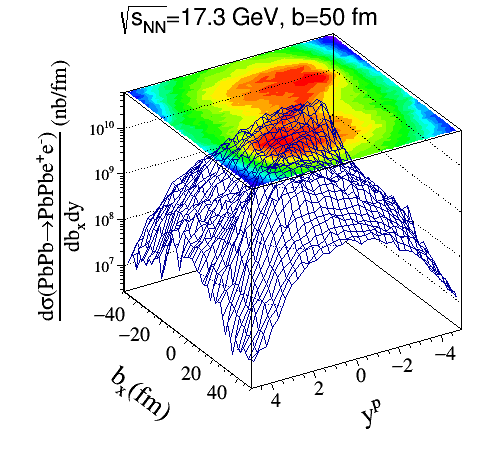}}\vspace{-1.0cm}\\
{\bf\hspace{1.5cm} (a) \hspace{5.0cm} (b)}\\
\vspace{0.5cm}
\resizebox{0.99\textwidth}{!}{%
\includegraphics[scale=0.4]{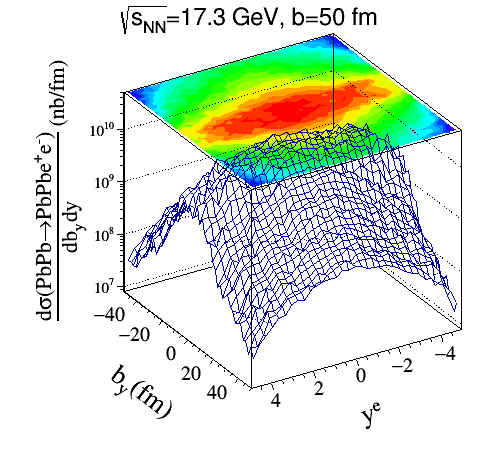}
\includegraphics[scale=0.4]{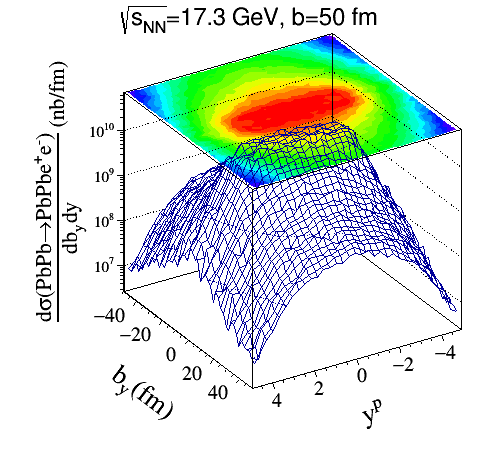}}\vspace{-1.0cm}\\
{\bf\hspace{1.5cm} (c) \hspace{5.0cm} (d)}\\
\caption{ Distribution of electrons ((a), (c)) and positrons ((b), (d)) for $\sqrt{s_{NN}}$= 17.3
 ~GeV at b=50~fm 
     integrated over ($b_x,b_y$)=(-100~fm, 100~fm), 
   $p_T^{ini}$=(0, 0.1~GeV).}
     \label{fig09}
\end{figure}
\begin{figure}[!hbt]
\resizebox{0.99\textwidth}{!}{%
\includegraphics[scale=0.4]{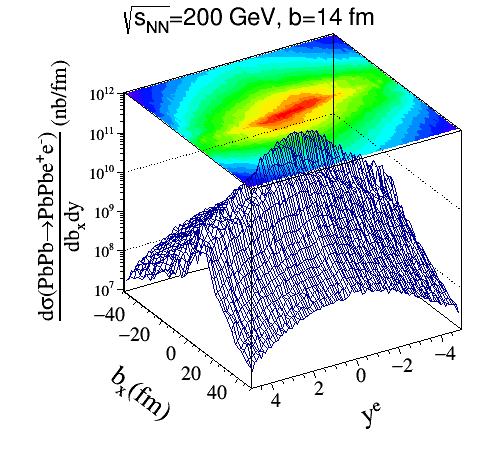}
\includegraphics[scale=0.4]{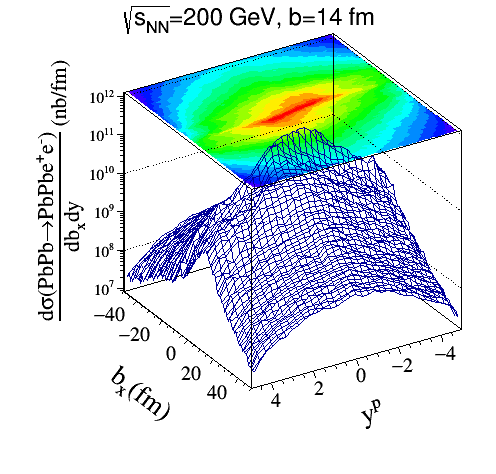}}\vspace{-1.0cm}\\
{\bf\hspace{1.5cm} (a) \hspace{5.0cm} (b)}\\
\resizebox{0.99\textwidth}{!}{%
\includegraphics[scale=0.4]{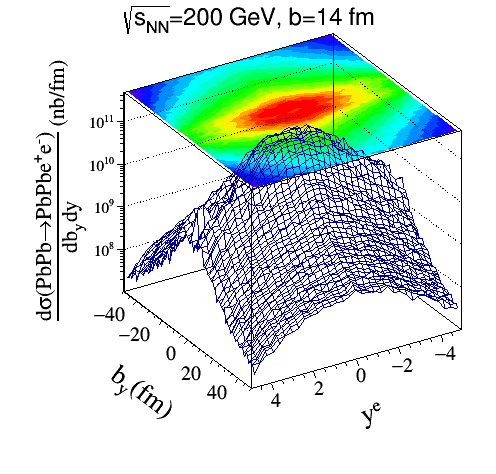}
\includegraphics[scale=0.4]{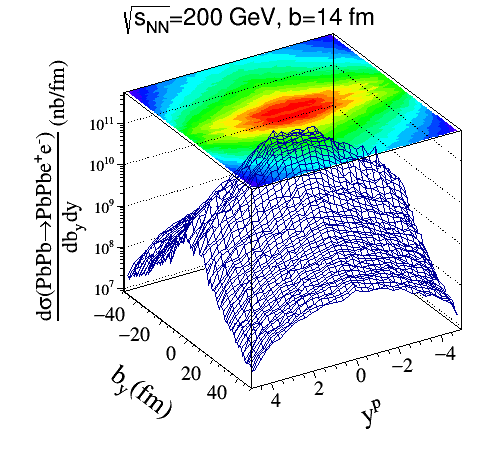}}\vspace{-1.0cm}\\
{\bf\hspace{1.5cm} (c) \hspace{5.0cm} (d)}\\
\vspace{0.5cm}
\caption{ Distribution of electrons ((a), (c)) and positrons ((b), (d)) for $\sqrt{s_{NN}}$= 200
 ~GeV at b=14~fm 
integrated over ($b_x,b_y$)=(-50~fm, 50~fm) for $p_T^{ini}$=(0, 0.1~GeV)}
     \label{fig11}
\end{figure}

The dependence on the position of emission and final rapidity allows to 
understand how the geometry influences the electromagnetic interaction between 
leptons and nuclei. Figures \ref{fig08},\ref{fig09} 
and \ref{fig11} present the cross section distribution in  
$b_x$ (top rows) and rapidity for electrons (a) and positrons (b) and (bottom rows)
$b_y$ and rapidity for electrons (c) and positrons (d).
Figs.~\ref{fig08} and \ref{fig09} are for $\sqrt{s_{NN}}$= 17.3~GeV but with 
the impact parameter 14$\pm$0.05 and 50$\pm$0.05~fm.
Fig.~\ref{fig11} is for $\sqrt{s_{NN}}$= 200~GeV and
b=14~fm. 
These plots allow to investigate the anisotropy caused by the
interaction between leptons and nuclei. It is more visible for larger 
impact parameter when the spectators are well separated (Fig.~\ref{fig09}). 
\begin{figure*}
\resizebox{0.99\textwidth}{!}{%
\includegraphics{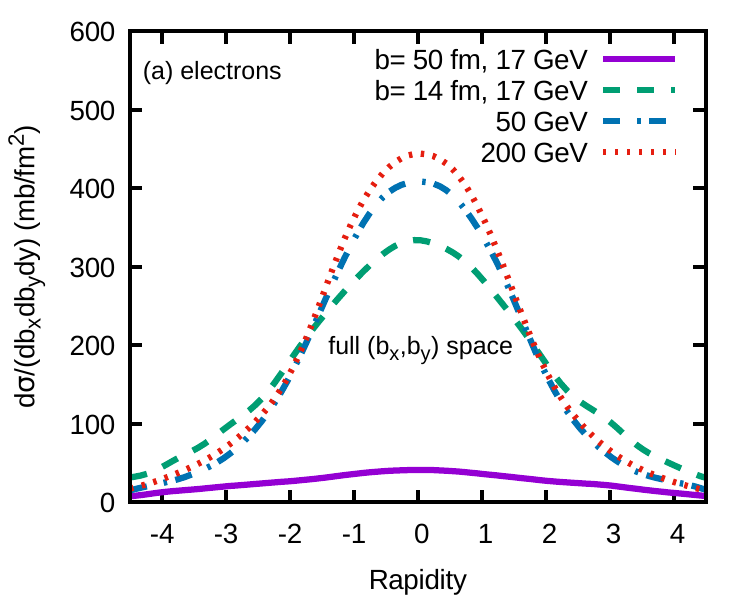}
\includegraphics{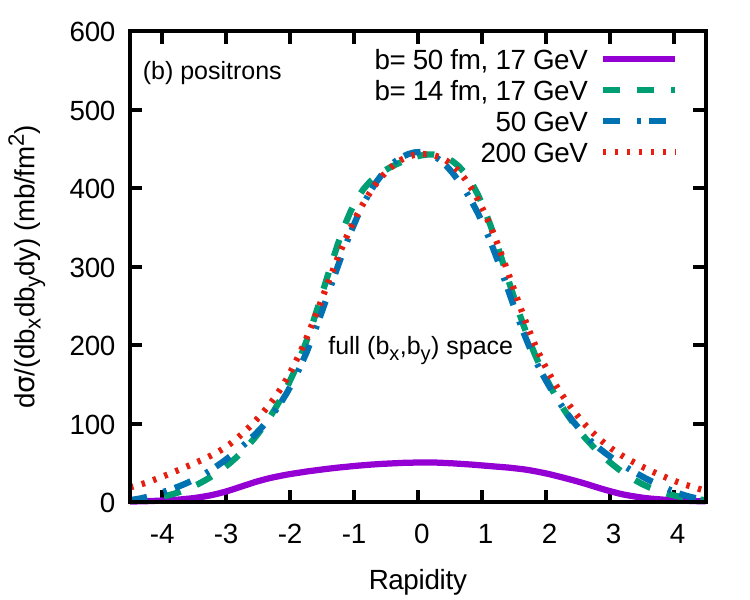}}
\caption{Rapidity distribution of electrons for $\sqrt{s_{NN}}$= 17.3~GeV (b=14~fm,
  50~fm) and 50~GeV and 200~GeV with b=14$\pm$0.05~fm (only) integrated over 
($b_x,b_y$)=(-50~fm, 50~fm), $p_T^{ini}$=(0, 0.1~GeV).}
     \label{fig13}
\end{figure*}

\begin{figure}
\resizebox{0.49\textwidth}{!}{%
\includegraphics{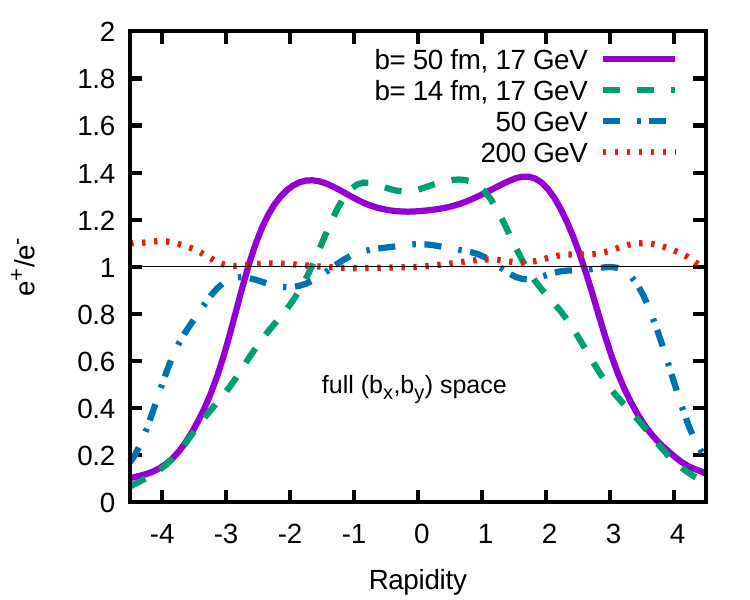}}
\caption{The ratio of rapidity distributions of positrons and electrons for 
  $\sqrt{s_{NN}}$= 17.3~GeV and fixed b=14~fm and 50~fm 
  and $\sqrt{s_{NN}}$ = 50~GeV and 200~GeV with fixed b=14~fm 
  integrated over ($b_x,b_y$)=(-50~fm, 50~fm) and transverse momenta in the interval 
   $p_T^{ini}$=(0, 0.1~GeV).}
     \label{fig14}
\end{figure}

Distributions for fixed impact parameter and different beam energies 
reflect the behavior seen in Fig.~\ref{fig05b}. For collision with 
$\sqrt{s_{NN}}$= 200~GeV (Fig.~\ref{fig11})
the electrons and positrons almost do not feel the presence of the EM
fields of the nuclei.  

Integrating over full ($b_x,b_y$) plane one obtains 
the rapidity distribution of final leptons
shown in Fig.~\ref{fig13} (a) separately for electrons and (b) positrons, 
for two impact parameters: 14~fm (full lines) and 50~fm (dashed lines). 
Electrons have a somewhat wider distribution than positrons and 
this is independent of the impact parameter.
While positron rapidity distributions only weakly depend on collision energy it is not the case for electrons, where sizeable differences can be observed.

The ratio of distributions for positrons and electrons (Fig.~\ref{fig14}) reflects the behavior seen in Figs.~\ref{fig08} and \ref{fig09}. 
This plot shows a combined effect of EPA cross section and 
the EM interactions of leptons with nuclei. Thus despite the production 
cross section is lower for larger impact parameter, the discussed
phenomenon should be visible for larger rapidities. 
The ratio quickly changes with energy and tends to 1 for larger energies (see dotted line for $\sqrt{{s}_{NN}}$ = 200~GeV).

Finally, we wish to make a supplementary comment on applicability
	of the classical approach used in the present paper.
	In principle, one could worry about uncertainty principle.
	In our calculations we assume that we know simultaneously momenta
	and positions of electrons/positrons.
	In the region of interest momenta of electrons/positrons are not small.
	Only transverse momenta are small.
	As shown in our paper, for the EM interactions the exact position 
	is not crucial for final distributions. The effect becomes smaller only when
	the initial distances are dramatically modified. So we feel 
	the uncertainty principle is not a crucial limitation for our study.
	In addition, we showed in our earlier calculations for
	pion production \cite{Rybicki:2006qm,Ozvenchuk:2019cve} that the dramatic effects for pion velocities close to
	the velocity of one of the heavy ion beams can be well described
	by the classical approach.

\section{Conclusions}

In the last 15 years the electromagnetic effects due to a large (moving)
charge of the spectators on charged pion momentum distributions were 
observed both in theoretical calculations and experimentally 
in peripheral heavy ion collisions at SPS and RHIC energies.
Interesting and sometimes spectacular effects were identified.

In the present paper we have discussed whether such effects could also
be observed for the distributions of electrons/positrons produced
via photon-photon fusion in heavy ion UPCs.
The corresponding cross section can be rather reliably calculated
and turned out to be large, especially for low transverse
momentum electrons/positrons. The impact parameter equivalent photon
approximation is well suitable for investigating the electromagnetic
effects.
On the experimental side only rather large transverse momentum
electrons/positrons could be measured so far at RHIC and the LHC, 
typically larger than 0.5~GeV.

We have organized calculations that include the EM effects using as an input EPA distributions.
First multi-differential (in momenta and impact parameter) distributions 
for the diphoton production of the $e^+ e^-$ pairs are prepared in 
the impact parameter equivalent photon approximation. 
Such distributions are used next to calculate the propagation of 
the electrons/positrons in strong EM fields generated by 
the quickly moving nuclei.
The propagation has been done by solving numerically relativistic 
equation of motion. Strong EM effects have been observed only at 
very small transverse momenta of electron/positron. Therefore 
to accelerate calculations we have limited to really small initial 
transverse momenta $p_T <$ 0.1~GeV.

The shape of the differential cross section in rapidity and transverse
momentum does not depend on the energy of the process but rather 
on the emission point in the impact parameter plane ($b_x,b_y$). 
We have investigated effects for different initial conditions, 
i.e. different emission positions in the impact parameter space.

The leptons interact electromagnetically with charged nuclei which 
changes their trajectories. The biggest effect has been identified for 
the CM emission point. 
However, the integration over full ($b_x,b_y$) plane washes out this effect to large extent. 
The range of $p_T$=(0, 0.02~GeV) has turned out the most preferable 
to investigate the influence of the EM effects on leptons originating from various 
($b_x,b_y$) plane points.

Moreover, the impact parameter influences not only the value of 
the cross section but also the shapes of distributions of final leptons. 

The $AA \rightarrow AAe^+e^-$ process creates leptons in a broad
range of rapidities. We have found that only at small transverse momenta of
electrons/positrons one can observe sizeable EM effects.
 
The performed calculations allow to conclude that the maximal 
beam energy for the Pb+Pb collision, where the EM effects between leptons
and nuclei are evident at midrapidities, is probably 
$\sqrt{{s}_{NN}}$=100~GeV.
Observation of this outcome at higher energies may be therefore 
rather difficult, if not impossible.
The effect survives even at high energies but close to beam rapidity. 
However, this region of the phase space is usually not instrumented
and does not allow electron/positron measurement.

So far the effect of EM interaction was studied for fixed values of 
the impact parameter (mainly for $b$ = 14~fm). However, the impact parameter
cannot be measured.
The integration over the impact parameter is rather difficult and
goes beyond the scope of the present paper.
Such an integration will be studied elsewhere.

Our exploratory calculations have been done using impact parameter EPA.
The discussed here final state electromagnetic effects were found at relatively small transverse momenta where more refined approach, including impact parameter-momentum correlations, may be necessary.
Here we have discussed the differences of the approach only for the production cross-section. Combining the FSI results and impact parameter - momentum correlations is difficult but could be done provided there is a chance to see the FSI effects experimentally.

On the experimental side a good measurement of electrons and positrons
at low transverse momenta ($p_T <$ 0.1~GeV) is necessary to see 
an effect.
In principle, the measurement of the $e^+ / e^-$ ratio as a function
of lepton rapidity and transverse momentum would be useful in 
this context.
According to our knowledge there are definite plans only for
high energies (ALICE-3 project) where, however, the EM effect should
be very small (high collision energy). At CERN SPS energies the effect
is rather large but at very small transverse momenta.
RHIC would probably be a good place to observe
the effect of the discussed here EM interactions but this would
require a modification of the present apparatus.

The discussed here effects could be interesting in future at the AFTER$@$LHC experiment \cite{AFTER,AFTER2},
where $\sqrt{s_{NN}} \sim 100$~GeV.
It may be also interesting to consider in future the problem in a quantal/field-theoretical approaches,
but this is not a easy task. 

{\bf Acknowledgement}

A.S. is indebted to Andrzej Rybicki for past collaboration on
electromagnetic effects in heavy ion collisions and Wolfgang Sch\"afer for a discussion on quantal effects.
This work is partially supported by
the Polish National Science Centre under Grant
No. 2018/31/B/ST2/03537
and by the Center for Innovation and Transfer of Natural Sciences 
and Engineering Knowledge in Rzesz\'ow (Poland).


%

\bibliography{biblio}


\begin{thebibliography}{58}
\ifx \bisbn   \undefined \def \bisbn  #1{ISBN #1}\fi
\ifx \binits  \undefined \def \binits#1{#1}\fi
\ifx \bauthor  \undefined \def \bauthor#1{#1}\fi
\ifx \batitle  \undefined \def \batitle#1{#1}\fi
\ifx \bjtitle  \undefined \def \bjtitle#1{#1}\fi
\ifx \bvolume  \undefined \def \bvolume#1{\textbf{#1}}\fi
\ifx \byear  \undefined \def \byear#1{#1}\fi
\ifx \bissue  \undefined \def \bissue#1{#1}\fi
\ifx \bfpage  \undefined \def \bfpage#1{#1}\fi
\ifx \blpage  \undefined \def \blpage #1{#1}\fi
\ifx \burl  \undefined \def \burl#1{\textsf{#1}}\fi
\ifx \doiurl  \undefined \def \doiurl#1{\url{https://doi.org/#1}}\fi
\ifx \betal  \undefined \def \betal{\textit{et al.}}\fi
\ifx \binstitute  \undefined \def \binstitute#1{#1}\fi
\ifx \binstitutionaled  \undefined \def \binstitutionaled#1{#1}\fi
\ifx \bctitle  \undefined \def \bctitle#1{#1}\fi
\ifx \beditor  \undefined \def \beditor#1{#1}\fi
\ifx \bpublisher  \undefined \def \bpublisher#1{#1}\fi
\ifx \bbtitle  \undefined \def \bbtitle#1{#1}\fi
\ifx \bedition  \undefined \def \bedition#1{#1}\fi
\ifx \bseriesno  \undefined \def \bseriesno#1{#1}\fi
\ifx \blocation  \undefined \def \blocation#1{#1}\fi
\ifx \bsertitle  \undefined \def \bsertitle#1{#1}\fi
\ifx \bsnm \undefined \def \bsnm#1{#1}\fi
\ifx \bsuffix \undefined \def \bsuffix#1{#1}\fi
\ifx \bparticle \undefined \def \bparticle#1{#1}\fi
\ifx \barticle \undefined \def \barticle#1{#1}\fi
\bibcommenthead
\ifx \bconfdate \undefined \def \bconfdate #1{#1}\fi
\ifx \botherref \undefined \def \botherref #1{#1}\fi
\ifx \url \undefined \def \url#1{\textsf{#1}}\fi
\ifx \bchapter \undefined \def \bchapter#1{#1}\fi
\ifx \bbook \undefined \def \bbook#1{#1}\fi
\ifx \bcomment \undefined \def \bcomment#1{#1}\fi
\ifx \oauthor \undefined \def \oauthor#1{#1}\fi
\ifx \citeauthoryear \undefined \def \citeauthoryear#1{#1}\fi
\ifx \endbibitem  \undefined \def \endbibitem {}\fi
\ifx \bconflocation  \undefined \def \bconflocation#1{#1}\fi
\ifx \arxivurl  \undefined \def \arxivurl#1{\textsf{#1}}\fi
\csname PreBibitemsHook\endcsname

\bibitem{Rybicki:2006qm}
\begin{barticle}
\bauthor{\bsnm{Rybicki}, \binits{A.}},
\bauthor{\bsnm{Szczurek}, \binits{A.}}:
\batitle{{The Spectator Electromagnetic Effect on Charged Pion Spectra in
  Peripheral Ultrarelativistic Heavy-Ion Collisions}}.
\bjtitle{Phys. Rev. C}
\bvolume{75},
\bfpage{054903}
(\byear{2007})
{\href{https://arxiv.org/abs/nucl-th/0610036}{{arXiv:nucl-th/0610036}}}.
\doiurl{10.1103/PhysRevC.75.054903}
\end{barticle}
\endbibitem

\bibitem{Rybicki:2013qla}
\begin{barticle}
\bauthor{\bsnm{Rybicki}, \binits{A.}},
\bauthor{\bsnm{Szczurek}, \binits{A.}}:
\batitle{{Spectator induced electromagnetic effect on directed flow in heavy
  ion collisions}}.
\bjtitle{Phys. Rev. C}
\bvolume{87}(\bissue{5}),
\bfpage{054909}
(\byear{2013})
{\href{https://arxiv.org/abs/1303.7354}{{arXiv:1303.7354}}}
{[nucl-th]}.
\doiurl{10.1103/PhysRevC.87.054909}
\end{barticle}
\endbibitem

\bibitem{Rybicki:2009zz}
\begin{barticle}
\bauthor{\bsnm{Rybicki}, \binits{A.}}:
\batitle{{Strong and electromagnetic interactions at SPS energies}}.
\bjtitle{PoS}
\bvolume{EPS-HEP2009},
\bfpage{031}
(\byear{2009}).
\doiurl{10.22323/1.084.0031}
\end{barticle}
\endbibitem

\bibitem{Schlagheck:1999aq}
\begin{barticle}
\bauthor{\bsnm{Schlagheck}, \binits{H.}}:
\batitle{{Thermalization and flow in 158-GeV/A Pb + Pb collisions}}.
\bjtitle{Nucl. Phys. A}
\bvolume{663},
\bfpage{725}--\blpage{728}
(\byear{2000})
{\href{https://arxiv.org/abs/nucl-ex/9909005}{{arXiv:nucl-ex/9909005}}}.
\doiurl{10.1016/S0375-9474(99)00703-4}
\end{barticle}
\endbibitem

\bibitem{Rybicki:2014rna}
\begin{botherref}
\oauthor{\bsnm{Rybicki}, \binits{A.}},
\oauthor{\bsnm{Szczurek}, \binits{A.}}:
{Charge splitting of directed flow and space-time picture of pion emission from
  the electromagnetic interactions with spectators}
(2014)
{\href{https://arxiv.org/abs/{1405.6860}}{{{arXiv}:{1405.6860}}}}
{[nucl-th]}
\end{botherref}
\endbibitem

\bibitem{Ozvenchuk:2019cve}
\begin{barticle}
\bauthor{\bsnm{Ozvenchuk}, \binits{V.}},
\bauthor{\bsnm{Rybicki}, \binits{A.}},
\bauthor{\bsnm{Szczurek}, \binits{A.}},
\bauthor{\bsnm{Marcinek}, \binits{A.}},
\bauthor{\bsnm{Kie\l{}bowicz}, \binits{M.}}:
\batitle{{Spectator induced electromagnetic effects in heavy-ion collisions and
  space-time-momentum conditions for pion emission}}.
\bjtitle{Phys. Rev. C}
\bvolume{102}(\bissue{1}),
\bfpage{014901}
(\byear{2020})
{\href{https://arxiv.org/abs/1910.04544}{{arXiv:1910.04544}}}
{[nucl-th]}.
\doiurl{10.1103/PhysRevC.102.014901}
\end{barticle}
\endbibitem

\bibitem{Klusek-Gawenda:2016suk}
\begin{barticle}
\bauthor{\bsnm{K\l{}usek-Gawenda}, \binits{M.}},
\bauthor{\bsnm{Szczurek}, \binits{A.}}:
\batitle{{Double scattering production of two positron\textendash{}electron
  pairs in ultraperipheral heavy-ion collisions}}.
\bjtitle{Phys. Lett. B}
\bvolume{763},
\bfpage{416}--\blpage{421}
(\byear{2016})
{\href{https://arxiv.org/abs/1607.05095}{{arXiv:1607.05095}}}
{[nucl-th]}.
\doiurl{10.1016/j.physletb.2016.10.079}
\end{barticle}
\endbibitem

\bibitem{vanHameren:2017krz}
\begin{barticle}
\bauthor{\bparticle{van} \bsnm{Hameren}, \binits{A.}},
\bauthor{\bsnm{K\l{}usek-Gawenda}, \binits{M.}},
\bauthor{\bsnm{Szczurek}, \binits{A.}}:
\batitle{{Single- and double-scattering production of four muons in
  ultraperipheral $PbPb$ collisions at the Large Hadron Collider}}.
\bjtitle{Phys. Lett. B}
\bvolume{776},
\bfpage{84}--\blpage{90}
(\byear{2018})
{\href{https://arxiv.org/abs/1708.07742}{{arXiv:1708.07742}}}
{[hep-ph]}.
\doiurl{10.1016/j.physletb.2017.11.029}
\end{barticle}
\endbibitem

\bibitem{Dirac:1930}
\begin{barticle}
\bauthor{\bsnm{Dirac}, \binits{P.A.M.}}:
\batitle{On the annihilation of electrons and protons}.
\bjtitle{Mathematical Proceedings of the Cambridge Philosophical Society}
\bvolume{26}(\bissue{3}),
\bfpage{361}--\blpage{375}
(\byear{1930}).
\doiurl{10.1017/S0305004100016091}
\end{barticle}
\endbibitem

\bibitem{Anderson}
\begin{barticle}
\bauthor{\bsnm{Anderson}, \binits{C.D.}}:
\batitle{The positive electron}.
\bjtitle{Phys. Rev.}
\bvolume{43},
\bfpage{491}--\blpage{494}
(\byear{1933}).
\doiurl{10.1103/PhysRev.43.491}
\end{barticle}
\endbibitem

\bibitem{Breit:1934zz}
\begin{barticle}
\bauthor{\bsnm{Breit}, \binits{G.}},
\bauthor{\bsnm{Wheeler}, \binits{J.A.}}:
\batitle{{Collision of two light quanta}}.
\bjtitle{Phys. Rev.}
\bvolume{46}(\bissue{12}),
\bfpage{1087}--\blpage{1091}
(\byear{1934}).
\doiurl{10.1103/PhysRev.46.1087}
\end{barticle}
\endbibitem

\bibitem{BH1934}
\begin{barticle}
\bauthor{\bsnm{Bethe}, \binits{H.A.}},
\bauthor{\bsnm{Heitler}, \binits{W.}}:
\batitle{{On the stopping of fast particles and on the creation of positive
  electrons}}.
\bjtitle{Proc. R. Soc. Lond. A}
\bvolume{146},
\bfpage{83}--\blpage{112}
(\byear{1934}).
\doiurl{10.1098/rspa.1934.0140}
\end{barticle}
\endbibitem

\bibitem{LL1934}
\begin{barticle}
\bauthor{\bsnm{Landau}, \binits{L.D.}},
\bauthor{\bsnm{Lifshitz}, \binits{E.}}:
\batitle{{On the production of electrons and positrons by a collision of two
  particles}}.
\bjtitle{Phys. Z. Sowjet.}
\bvolume{6},
\bfpage{244}
(\byear{1934}).
\doiurl{10.1103/PhysRevA.69.022708}
\end{barticle}
\endbibitem

\bibitem{Williams:1935}
\begin{barticle}
\bauthor{\bsnm{Williams}, \binits{E.}}:
\batitle{{Production of Electron-Positron Pairs}}.
\bjtitle{Nature}
\bvolume{135},
\bfpage{66}
(\byear{1935}).
\doiurl{10.1038/135066a0}
\end{barticle}
\endbibitem

\bibitem{Hubbel:2006}
\begin{barticle}
\bauthor{\bsnm{Hubbell}, \binits{J.H.}}:
\batitle{{Electron–positron pair production by photons: A historical
  overview}}.
\bjtitle{Radiation Physics and Chemistry}
\bvolume{75}(\bissue{6}),
\bfpage{614}--\blpage{623}
(\byear{2006}).
\doiurl{10.1016/j.radphyschem.2005.10.008}.
\bcomment{Pair Production}
\end{barticle}
\endbibitem

\bibitem{Ruffini:2009hg}
\begin{barticle}
\bauthor{\bsnm{Ruffini}, \binits{R.}},
\bauthor{\bsnm{Vereshchagin}, \binits{G.}},
\bauthor{\bsnm{Xue}, \binits{S.-S.}}:
\batitle{{Electron-positron pairs in physics and astrophysics: from heavy
  nuclei to black holes}}.
\bjtitle{Phys. Rept.}
\bvolume{487},
\bfpage{1}--\blpage{140}
(\byear{2010})
{\href{https://arxiv.org/abs/0910.0974}{{arXiv:0910.0974}}}
{[astro-ph.HE]}.
\doiurl{10.1016/j.physrep.2009.10.004}
\end{barticle}
\endbibitem

\bibitem{Klusek-Gawenda:2010vqb}
\begin{barticle}
\bauthor{\bsnm{K\l{}usek-Gawenda}, \binits{M.}},
\bauthor{\bsnm{Szczurek}, \binits{A.}}:
\batitle{{Exclusive muon-pair productions in ultrarelativistic heavy-ion
  collisions -- realistic nucleus charge form factor and differential
  distributions}}.
\bjtitle{Phys. Rev. C}
\bvolume{82},
\bfpage{014904}
(\byear{2010})
{\href{https://arxiv.org/abs/1004.5521}{{arXiv:1004.5521}}}
{[nucl-th]}.
\doiurl{10.1103/PhysRevC.82.014904}
\end{barticle}
\endbibitem

\bibitem{Klusek-Gawenda:2013ema}
\begin{barticle}
\bauthor{\bsnm{K\l{}usek-Gawenda}, \binits{M.}},
\bauthor{\bsnm{Ciema\l{}a}, \binits{M.}},
\bauthor{\bsnm{Sch\"afer}, \binits{W.}},
\bauthor{\bsnm{Szczurek}, \binits{A.}}:
\batitle{{Electromagnetic excitation of nuclei and neutron evaporation in
  ultrarelativistic ultraperipheral heavy ion collisions}}.
\bjtitle{Phys. Rev. C}
\bvolume{89}(\bissue{5}),
\bfpage{054907}
(\byear{2014})
{\href{https://arxiv.org/abs/1311.1938}{{arXiv:1311.1938}}}
{[nucl-th]}.
\doiurl{10.1103/PhysRevC.89.054907}
\end{barticle}
\endbibitem

\bibitem{STAR1}
\begin{barticle}
\bauthor{\bsnm{Adams}, \binits{J.}}, \betal:
\batitle{{Production of e+ e- pairs accompanied by nuclear dissociation in
  ultra-peripheral heavy ion collision}}.
\bjtitle{Phys. Rev. C}
\bvolume{70},
\bfpage{031902}
(\byear{2004})
{\href{https://arxiv.org/abs/nucl-ex/0404012}{{arXiv:nucl-ex/0404012}}}.
\doiurl{10.1103/PhysRevC.70.031902}
\end{barticle}
\endbibitem

\bibitem{STAR2}
\begin{barticle}
\bauthor{\bsnm{Adam}, \binits{J.}}, \betal:
\batitle{{Low-$p_T$ $e^{+}e^{-}$ pair production in Au$+$Au collisions at
  $\sqrt{s_{NN}}$ = 200 GeV and U$+$U collisions at $\sqrt{s_{NN}}$ = 193 GeV
  at STAR}}.
\bjtitle{Phys. Rev. Lett.}
\bvolume{121}(\bissue{13}),
\bfpage{132301}
(\byear{2018})
{\href{https://arxiv.org/abs/1806.02295}{{arXiv:1806.02295}}}
{[hep-ex]}.
\doiurl{10.1103/PhysRevLett.121.132301}
\end{barticle}
\endbibitem

\bibitem{ATLAS1}
\begin{barticle}
\bauthor{\bsnm{Aaboud}, \binits{M.}}, \betal:
\batitle{{Observation of centrality-dependent acoplanarity for muon pairs
  produced via two-photon scattering in Pb+Pb collisions at
  $\sqrt{s_{\mathrm{NN}}}=5.02$ TeV with the ATLAS detector}}.
\bjtitle{Phys. Rev. Lett.}
\bvolume{121}(\bissue{21}),
\bfpage{212301}
(\byear{2018})
{\href{https://arxiv.org/abs/1806.08708}{{arXiv:1806.08708}}}
{[nucl-ex]}.
\doiurl{10.1103/PhysRevLett.121.212301}
\end{barticle}
\endbibitem

\bibitem{ATLAS2}
\begin{barticle}
\bauthor{\bsnm{Aad}, \binits{G.}}, \betal:
\batitle{{Exclusive dimuon production in ultraperipheral Pb+Pb collisions at
  $\sqrt{s_{\mathrm{NN}}} = 5.02$ TeV with ATLAS}}.
\bjtitle{Phys. Rev. C}
\bvolume{104},
\bfpage{024906}
(\byear{2021})
{\href{https://arxiv.org/abs/2011.12211}{{arXiv:2011.12211}}}
{[nucl-ex]}.
\doiurl{10.1103/PhysRevC.104.024906}
\end{barticle}
\endbibitem

\bibitem{k_t-factorization}
\begin{barticle}
\bauthor{\bsnm{K\l{}usek-Gawenda}, \binits{M.}},
\bauthor{\bsnm{Rapp}, \binits{R.}},
\bauthor{\bsnm{Sch\"afer}, \binits{W.}},
\bauthor{\bsnm{Szczurek}, \binits{A.}}:
\batitle{{Dilepton Radiation in Heavy-Ion Collisions at Small Transverse
  Momentum}}.
\bjtitle{Phys. Lett. B}
\bvolume{790},
\bfpage{339}--\blpage{344}
(\byear{2019})
{\href{https://arxiv.org/abs/1809.07049}{{arXiv:1809.07049}}}
{[nucl-th]}.
\doiurl{10.1016/j.physletb.2019.01.035}
\end{barticle}
\endbibitem

\bibitem{LZZ2019}
\begin{barticle}
\bauthor{\bsnm{Li}, \binits{C.}},
\bauthor{\bsnm{Zhou}, \binits{J.}},
\bauthor{\bsnm{Zhou}, \binits{Y.-J.}}:
\batitle{{Probing the linear polarization of photons in ultraperipheral heavy
  ion collisions}}.
\bjtitle{Phys. Lett. B}
\bvolume{795},
\bfpage{576}--\blpage{580}
(\byear{2019})
{\href{https://arxiv.org/abs/1903.10084}{{arXiv:1903.10084}}}
{[hep-ph]}.
\doiurl{10.1016/j.physletb.2019.07.005}
\end{barticle}
\endbibitem

\bibitem{ZBTX2020}
\begin{barticle}
\bauthor{\bsnm{Zha}, \binits{W.}},
\bauthor{\bsnm{Brandenburg}, \binits{J.D.}},
\bauthor{\bsnm{Tang}, \binits{Z.}},
\bauthor{\bsnm{Xu}, \binits{Z.}}:
\batitle{{Initial transverse-momentum broadening of Breit-Wheeler process in
  relativistic heavy-ion collisions}}.
\bjtitle{Phys. Lett. B}
\bvolume{800},
\bfpage{135089}
(\byear{2020})
{\href{https://arxiv.org/abs/1812.02820}{{arXiv:1812.02820}}}
{[nucl-th]}.
\doiurl{10.1016/j.physletb.2019.135089}
\end{barticle}
\endbibitem

\bibitem{KMXY2020}
\begin{barticle}
\bauthor{\bsnm{Klein}, \binits{S.}},
\bauthor{\bsnm{Mueller}, \binits{A.H.}},
\bauthor{\bsnm{Xiao}, \binits{B.-W.}},
\bauthor{\bsnm{Yuan}, \binits{F.}}:
\batitle{{Lepton Pair Production Through Two Photon Process in Heavy Ion
  Collisions}}.
\bjtitle{Phys. Rev. D}
\bvolume{102}(\bissue{9}),
\bfpage{094013}
(\byear{2020})
{\href{https://arxiv.org/abs/2003.02947}{{arXiv:2003.02947}}}
{[hep-ph]}.
\doiurl{10.1103/PhysRevD.102.094013}
\end{barticle}
\endbibitem

\bibitem{WPW2021}
\begin{barticle}
\bauthor{\bsnm{Wang}, \binits{R.-j.}},
\bauthor{\bsnm{Pu}, \binits{S.}},
\bauthor{\bsnm{Wang}, \binits{Q.}}:
\batitle{{Lepton pair production in ultraperipheral collisions}}.
\bjtitle{Phys. Rev. D}
\bvolume{104}(\bissue{5}),
\bfpage{056011}
(\byear{2021})
{\href{https://arxiv.org/abs/2106.05462}{{arXiv:2106.05462}}}
{[hep-ph]}.
\doiurl{10.1103/PhysRevD.104.056011}
\end{barticle}
\endbibitem

\bibitem{Wigner}
\begin{barticle}
\bauthor{\bsnm{K\l{}usek-Gawenda}, \binits{M.}},
\bauthor{\bsnm{Sch\"afer}, \binits{W.}},
\bauthor{\bsnm{Szczurek}, \binits{A.}}:
\batitle{{Centrality dependence of dilepton production via $\gamma \gamma$
  processes from Wigner distributions of photons in nuclei}}.
\bjtitle{Phys. Lett. B}
\bvolume{814},
\bfpage{136114}
(\byear{2021})
{\href{https://arxiv.org/abs/2012.11973}{{arXiv:2012.11973}}}
{[hep-ph]}.
\doiurl{10.1016/j.physletb.2021.136114}
\end{barticle}
\endbibitem

\bibitem{Klusek-Gawenda:2018zfz}
\begin{barticle}
\bauthor{\bsnm{K\l{}usek-Gawenda}, \binits{M.}},
\bauthor{\bsnm{Rapp}, \binits{R.}},
\bauthor{\bsnm{Sch\"afer}, \binits{W.}},
\bauthor{\bsnm{Szczurek}, \binits{A.}}:
\batitle{{Dilepton Radiation in Heavy-Ion Collisions at Small Transverse
  Momentum}}.
\bjtitle{Phys. Lett. B}
\bvolume{790},
\bfpage{339}--\blpage{344}
(\byear{2019})
{\href{https://arxiv.org/abs/1809.07049}{{arXiv:1809.07049}}}
{[nucl-th]}.
\doiurl{10.1016/j.physletb.2019.01.035}
\end{barticle}
\endbibitem

\bibitem{Klusek-Gawenda:2020eja}
\begin{barticle}
\bauthor{\bsnm{K\l{}usek-Gawenda}, \binits{M.}},
\bauthor{\bsnm{Sch\"afer}, \binits{W.}},
\bauthor{\bsnm{Szczurek}, \binits{A.}}:
\batitle{{Centrality dependence of dilepton production via $\gamma \gamma$
  processes from Wigner distributions of photons in nuclei}}.
\bjtitle{Phys. Lett. B}
\bvolume{814},
\bfpage{136114}
(\byear{2021})
{\href{https://arxiv.org/abs/2012.11973}{{arXiv:2012.11973}}}
{[hep-ph]}.
\doiurl{10.1016/j.physletb.2021.136114}
\end{barticle}
\endbibitem

\bibitem{Rybicki:2011zz}
\begin{barticle}
\bauthor{\bsnm{Rybicki}, \binits{A.}}:
\batitle{{What Is the role of nuclear effects in ultrarelativistic reactions at
  158-GeV/nucleon?}}
\bjtitle{Acta Phys. Polon. B}
\bvolume{42},
\bfpage{867}--\blpage{876}
(\byear{2011}).
\doiurl{10.5506/APhysPolB.42.867}
\end{barticle}
\endbibitem

\bibitem{BM1954}
\begin{barticle}
\bauthor{\bsnm{Bethe}, \binits{H.A.}},
\bauthor{\bsnm{Maximon}, \binits{L.C.}}:
\batitle{{Theory of Bremsstrahlung and Pair Production. 1. Differential Cross
  Section}}.
\bjtitle{Phys. Rev.}
\bvolume{93},
\bfpage{768}--\blpage{784}
(\byear{1954}).
\doiurl{10.1103/PhysRev.93.768}
\end{barticle}
\endbibitem

\bibitem{IM1998}
\begin{barticle}
\bauthor{\bsnm{Ivanov}, \binits{D.}},
\bauthor{\bsnm{Melnikov}, \binits{K.}}:
\batitle{{Lepton pair production by a high-energy photon in a strong
  electromagnetic field}}.
\bjtitle{Phys. Rev. D}
\bvolume{57},
\bfpage{4025}--\blpage{4034}
(\byear{1998})
{\href{https://arxiv.org/abs/hep-ph/9709352}{{arXiv:hep-ph/9709352}}}.
\doiurl{10.1103/PhysRevD.57.4025}
\end{barticle}
\endbibitem

\bibitem{Tuchin2009}
\begin{barticle}
\bauthor{\bsnm{Tuchin}, \binits{K.}}:
\batitle{{Multi-photon interactions in lepton photo-production on nuclei at
  high energies}}.
\bjtitle{Phys. Rev. D}
\bvolume{80},
\bfpage{093006}
(\byear{2009})
{\href{https://arxiv.org/abs/0907.5189}{{arXiv:0907.5189}}}
{[hep-ph]}.
\doiurl{10.1103/PhysRevD.80.093006}
\end{barticle}
\endbibitem

\bibitem{SZZZ2020}
\begin{barticle}
\bauthor{\bsnm{Sun}, \binits{Z.-h.}},
\bauthor{\bsnm{Zheng}, \binits{D.-x.}},
\bauthor{\bsnm{Zhou}, \binits{J.}},
\bauthor{\bsnm{Zhou}, \binits{Y.-j.}}:
\batitle{{Studying Coulomb correction at EIC and EicC}}.
\bjtitle{Phys. Lett. B}
\bvolume{808},
\bfpage{135679}
(\byear{2020})
{\href{https://arxiv.org/abs/2002.07373}{{arXiv:2002.07373}}}
{[hep-ph]}.
\doiurl{10.1016/j.physletb.2020.135679}
\end{barticle}
\endbibitem

\bibitem{IKSS1998}
\begin{barticle}
\bauthor{\bsnm{Ivanov}, \binits{D.}},
\bauthor{\bsnm{Kuraev}, \binits{E.A.}},
\bauthor{\bsnm{Schiller}, \binits{A.}},
\bauthor{\bsnm{Serbo}, \binits{V.G.}}:
\batitle{{Production of e+ e- pairs to all orders in Z alpha for collisions of
  high-energy muons with heavy nuclei}}.
\bjtitle{Phys. Lett. B}
\bvolume{442},
\bfpage{453}--\blpage{458}
(\byear{1998})
{\href{https://arxiv.org/abs/hep-ph/9807311}{{arXiv:hep-ph/9807311}}}.
\doiurl{10.1016/S0370-2693(98)01278-7}
\end{barticle}
\endbibitem

\bibitem{LMS2004}
\begin{barticle}
\bauthor{\bsnm{Lee}, \binits{R.N.}},
\bauthor{\bsnm{Milstein}, \binits{A.I.}},
\bauthor{\bsnm{Strakhovenko}, \binits{V.M.}}:
\batitle{{High-energy expansion of Coulomb corrections to the e+ e-
  photoproduction cross-section}}.
\bjtitle{Phys. Rev. A}
\bvolume{69},
\bfpage{022708}
(\byear{2004})
{\href{https://arxiv.org/abs/hep-ph/0310108}{{arXiv:hep-ph/0310108}}}.
\doiurl{10.1103/PhysRevA.69.022708}
\end{barticle}
\endbibitem

\bibitem{SR2018}
\begin{botherref}
\oauthor{\bsnm{Sandrock}, \binits{A.}},
\oauthor{\bsnm{Rhode}, \binits{W.}}:
{Coulomb corrections to the bremsstrahlung and electron pair production cross
  section of high-energy muons on extended nuclei}
(2018)
{\href{https://arxiv.org/abs/{1807.08475}}{{{arXiv}:{1807.08475}}}}
{[{hep-ph}]}
\end{botherref}
\endbibitem

\bibitem{SW1998}
\begin{barticle}
\bauthor{\bsnm{Segev}, \binits{B.}},
\bauthor{\bsnm{Wells}, \binits{J.C.}}:
\batitle{{A Light fronts approach to electron - positron pair production in
  ultrarelativistic heavy ion collisions}}.
\bjtitle{Phys. Rev. A}
\bvolume{57},
\bfpage{1849}
(\byear{1998})
{\href{https://arxiv.org/abs/physics/9710008}{{arXiv:physics/9710008}}}.
\doiurl{10.1103/PhysRevA.57.1849}
\end{barticle}
\endbibitem

\bibitem{SW1999}
\begin{barticle}
\bauthor{\bsnm{Segev}, \binits{B.}},
\bauthor{\bsnm{Wells}, \binits{J.C.}}:
\batitle{{Exact Z**2 scaling of pair production in the high-energy limit of
  heavy ion collisions}}.
\bjtitle{Phys. Rev. C}
\bvolume{59},
\bfpage{2753}--\blpage{2756}
(\byear{1999})
{\href{https://arxiv.org/abs/physics/9805013}{{arXiv:physics/9805013}}}.
\doiurl{10.1103/PhysRevC.59.2753}
\end{barticle}
\endbibitem

\bibitem{ERSG1999}
\begin{barticle}
\bauthor{\bsnm{Eichmann}, \binits{U.}},
\bauthor{\bsnm{Reinhardt}, \binits{J.}},
\bauthor{\bsnm{Greiner}, \binits{W.}}:
\batitle{{Coulomb effects on electromagnetic pair production in
  ultrarelativistic heavy ion collisions}}.
\bjtitle{Phys. Rev. A}
\bvolume{59},
\bfpage{1223}--\blpage{1237}
(\byear{1999})
{\href{https://arxiv.org/abs/nucl-th/9806031}{{arXiv:nucl-th/9806031}}}.
\doiurl{10.1103/PhysRevA.59.1223}
\end{barticle}
\endbibitem

\bibitem{ISS1999}
\begin{barticle}
\bauthor{\bsnm{Ivanov}, \binits{D.Y.}},
\bauthor{\bsnm{Schiller}, \binits{A.}},
\bauthor{\bsnm{Serbo}, \binits{V.G.}}:
\batitle{{Large Coulomb corrections to the e+ e- pair production at
  relativistic heavy ion colliders}}.
\bjtitle{Phys. Lett. B}
\bvolume{454},
\bfpage{155}--\blpage{160}
(\byear{1999})
{\href{https://arxiv.org/abs/hep-ph/9809449}{{arXiv:hep-ph/9809449}}}.
\doiurl{10.1016/S0370-2693(99)00323-8}
\end{barticle}
\endbibitem

\bibitem{BGLP2001}
\begin{barticle}
\bauthor{\bsnm{Baltz}, \binits{A.J.}},
\bauthor{\bsnm{Gelis}, \binits{F.}},
\bauthor{\bsnm{McLerran}, \binits{L.D.}},
\bauthor{\bsnm{Peshier}, \binits{A.}}:
\batitle{{Coulomb corrections to e+ e- production in ultrarelativistic nuclear
  collisions}}.
\bjtitle{Nucl. Phys. A}
\bvolume{695},
\bfpage{395}--\blpage{429}
(\byear{2001})
{\href{https://arxiv.org/abs/nucl-th/0101024}{{arXiv:nucl-th/0101024}}}.
\doiurl{10.1016/S0375-9474(01)01109-5}
\end{barticle}
\endbibitem

\bibitem{LMS2002}
\begin{barticle}
\bauthor{\bsnm{Lee}, \binits{R.N.}},
\bauthor{\bsnm{Milstein}, \binits{A.I.}},
\bauthor{\bsnm{Serbo}, \binits{V.G.}}:
\batitle{{Structure of the Coulomb and unitarity corrections to the
  cross-section of e+ e- pair production in ultrarelativistic nuclear
  collisions}}.
\bjtitle{Phys. Rev. A}
\bvolume{65},
\bfpage{022102}
(\byear{2002})
{\href{https://arxiv.org/abs/hep-ph/0108014}{{arXiv:hep-ph/0108014}}}.
\doiurl{10.1103/PhysRevA.65.022102}
\end{barticle}
\endbibitem

\bibitem{ML1998}
\begin{barticle}
\bauthor{\bsnm{Baltz}, \binits{A.J.}},
\bauthor{\bsnm{McLerran}, \binits{L.D.}}:
\batitle{{Two center light cone calculation of pair production induced by
  ultrarelativistic heavy ions}}.
\bjtitle{Phys. Rev. C}
\bvolume{58},
\bfpage{1679}--\blpage{1688}
(\byear{1998})
{\href{https://arxiv.org/abs/nucl-th/9804042}{{arXiv:nucl-th/9804042}}}.
\doiurl{10.1103/PhysRevC.58.1679}
\end{barticle}
\endbibitem

\bibitem{B2003}
\begin{barticle}
\bauthor{\bsnm{Baltz}, \binits{A.J.}}:
\batitle{{Coulomb corrections in the calculation of ultrarelativistic heavy ion
  production of continuum e+ e- pairs}}.
\bjtitle{Phys. Rev. C}
\bvolume{68},
\bfpage{034906}
(\byear{2003})
{\href{https://arxiv.org/abs/nucl-th/0305083}{{arXiv:nucl-th/0305083}}}.
\doiurl{10.1103/PhysRevC.68.034906}
\end{barticle}
\endbibitem

\bibitem{BGKN2002}
\begin{barticle}
\bauthor{\bsnm{Bartos}, \binits{E.}},
\bauthor{\bsnm{Gevorkyan}, \binits{S.R.}},
\bauthor{\bsnm{Kuraev}, \binits{E.A.}},
\bauthor{\bsnm{Nikolaev}, \binits{N.N.}}:
\batitle{{Multiple lepton pair production in relativistic ion collisions}}.
\bjtitle{Phys. Lett. B}
\bvolume{538},
\bfpage{45}--\blpage{51}
(\byear{2002})
{\href{https://arxiv.org/abs/hep-ph/0204327}{{arXiv:hep-ph/0204327}}}.
\doiurl{10.1016/S0370-2693(02)01991-3}
\end{barticle}
\endbibitem

\bibitem{BGKN2004}
\begin{barticle}
\bauthor{\bsnm{Bartos}, \binits{E.}},
\bauthor{\bsnm{Gevorkyan}, \binits{S.R.}},
\bauthor{\bsnm{Kuraev}, \binits{E.A.}},
\bauthor{\bsnm{Nikolaev}, \binits{N.N.}}:
\batitle{{Multiple exchanges in lepton pair production in high-energy heavy ion
  collisions}}.
\bjtitle{J. Exp. Theor. Phys.}
\bvolume{100}(\bissue{4}),
\bfpage{645}--\blpage{655}
(\byear{2005})
{\href{https://arxiv.org/abs/hep-ph/0410263}{{arXiv:hep-ph/0410263}}}.
\doiurl{10.1134/1.1926426}
\end{barticle}
\endbibitem

\bibitem{BF90}
\begin{barticle}
\bauthor{\bsnm{Baur}, \binits{G.}},
\bauthor{\bsnm{Ferreira~Filho}, \binits{L.G.}}:
\batitle{{COHERENT PARTICLE PRODUCTION AT RELATIVISTIC HEAVY ION COLLIDERS
  INCLUDING STRONG ABSORPTION EFFECTS}}.
\bjtitle{Nucl. Phys. A}
\bvolume{518},
\bfpage{786}--\blpage{800}
(\byear{1990}).
\doiurl{10.1016/0375-9474(90)90191-N}
\end{barticle}
\endbibitem

\bibitem{Bertulani:2005ru}
\begin{barticle}
\bauthor{\bsnm{Bertulani}, \binits{C.A.}},
\bauthor{\bsnm{Klein}, \binits{S.R.}},
\bauthor{\bsnm{Nystrand}, \binits{J.}}:
\batitle{{Physics of ultra-peripheral nuclear collisions}}.
\bjtitle{Ann. Rev. Nucl. Part. Sci.}
\bvolume{55},
\bfpage{271}--\blpage{310}
(\byear{2005})
{\href{https://arxiv.org/abs/nucl-ex/0502005}{{arXiv:nucl-ex/0502005}}}.
\doiurl{10.1146/annurev.nucl.55.090704.151526}
\end{barticle}
\endbibitem

\bibitem{BGKN2009}
\begin{barticle}
\bauthor{\bsnm{Baltz}, \binits{A.J.}},
\bauthor{\bsnm{Gorbunov}, \binits{Y.}},
\bauthor{\bsnm{Klein}, \binits{S.R.}},
\bauthor{\bsnm{Nystrand}, \binits{J.}}:
\batitle{{Two-Photon Interactions with Nuclear Breakup in Relativistic Heavy
  Ion Collisions}}.
\bjtitle{Phys. Rev. C}
\bvolume{80},
\bfpage{044902}
(\byear{2009})
{\href{https://arxiv.org/abs/0907.1214}{{arXiv:0907.1214}}}
{[nucl-ex]}.
\doiurl{10.1103/PhysRevC.80.044902}
\end{barticle}
\endbibitem

\bibitem{STARlight}
\begin{barticle}
\bauthor{\bsnm{Klein}, \binits{S.R.}},
\bauthor{\bsnm{Nystrand}, \binits{J.}},
\bauthor{\bsnm{Seger}, \binits{J.}},
\bauthor{\bsnm{Gorbunov}, \binits{Y.}},
\bauthor{\bsnm{Butterworth}, \binits{J.}}:
\batitle{{STARlight: A Monte Carlo simulation program for ultra-peripheral
  collisions of relativistic ions}}.
\bjtitle{Comput. Phys. Commun.}
\bvolume{212},
\bfpage{258}--\blpage{268}
(\byear{2017})
{\href{https://arxiv.org/abs/1607.03838}{{arXiv:1607.03838}}}
{[hep-ph]}.
\doiurl{10.1016/j.cpc.2016.10.016}
\end{barticle}
\endbibitem

\bibitem{Vidovic1992}
\begin{barticle}
\bauthor{\bsnm{Vidovic}, \binits{M.}},
\bauthor{\bsnm{Greiner}, \binits{M.}},
\bauthor{\bsnm{Best}, \binits{C.}},
\bauthor{\bsnm{Soff}, \binits{G.}}:
\batitle{{Impact parameter dependence of the electromagnetic particle
  production in ultrarelativistic heavy ion collisions}}.
\bjtitle{Phys. Rev. C}
\bvolume{47},
\bfpage{2308}--\blpage{2319}
(\byear{1993}).
\doiurl{10.1103/PhysRevC.47.2308}
\end{barticle}
\endbibitem

\bibitem{HTB1995}
\begin{barticle}
\bauthor{\bsnm{Hencken}, \binits{K.}},
\bauthor{\bsnm{Trautmann}, \binits{D.}},
\bauthor{\bsnm{Baur}, \binits{G.}}:
\batitle{{Impact parameter dependence of the total probability for the
  electromagnetic electron - positron pair production in relativistic heavy ion
  collisions}}.
\bjtitle{Phys. Rev. A}
\bvolume{51},
\bfpage{1874}--\blpage{1882}
(\byear{1995})
{\href{https://arxiv.org/abs/nucl-th/9410014}{{arXiv:nucl-th/9410014}}}.
\doiurl{10.1103/PhysRevA.51.1874}
\end{barticle}
\endbibitem

\bibitem{ALICE}
\begin{barticle}
\bauthor{\bsnm{Abbas}, \binits{E.}}, \betal:
\batitle{{Charmonium and $e^+e^-$ pair photoproduction at mid-rapidity in
  ultra-peripheral Pb-Pb collisions at $\sqrt{s_{\rm NN}}$=2.76 TeV}}.
\bjtitle{Eur. Phys. J. C}
\bvolume{73}(\bissue{11}),
\bfpage{2617}
(\byear{2013})
{\href{https://arxiv.org/abs/1305.1467}{{arXiv:1305.1467}}}
{[nucl-ex]}.
\doiurl{10.1140/epjc/s10052-013-2617-1}
\end{barticle}
\endbibitem

\bibitem{Lepage:1977sw}
\begin{barticle}
\bauthor{\bsnm{Lepage}, \binits{G.P.}}:
\batitle{{A New Algorithm for Adaptive Multidimensional Integration}}.
\bjtitle{J. Comput. Phys.}
\bvolume{27},
\bfpage{192}
(\byear{1978}).
\doiurl{10.1016/0021-9991(78)90004-9}
\end{barticle}
\endbibitem

\bibitem{AFTER}
\begin{barticle}
\bauthor{\bsnm{Hadjidakis}, \binits{C.}}, \betal:
\batitle{{A fixed-target programme at the LHC: Physics case and projected
  performances for heavy-ion, hadron, spin and astroparticle studies}}.
\bjtitle{Phys. Rept.}
\bvolume{911},
\bfpage{1}--\blpage{83}
(\byear{2021})
{\href{https://arxiv.org/abs/1807.00603}{{arXiv:1807.00603}}}
{[hep-ex]}.
\doiurl{10.1016/j.physrep.2021.01.002}
\end{barticle}
\endbibitem

\bibitem{AFTER2}
\begin{botherref}
AFTER$@$LHC.
\url{http://after.in2p3.fr/}
\end{botherref}
\endbibitem

\end{thebibliography}

\end{document}